\documentclass[12pt]{iopart}
\pdfoutput=1
\expandafter\let\csname equation*\endcsname\relax
\expandafter\let\csname endequation*\endcsname\relax
\usepackage{amsfonts} 
\usepackage{amsmath}
\usepackage{iopams}
\usepackage{amssymb}
\usepackage{mathtools}
\usepackage{cancel} 
\usepackage{bm}
\usepackage[usenames, dvipsnames]{color} 
\usepackage{dcolumn}
\usepackage{epic} 
\usepackage{epsfig}
\usepackage{feynmp}
\usepackage{graphicx}
\usepackage{grffile}
\usepackage[breaklinks,colorlinks = true,linkcolor = red,urlcolor=blue,citecolor=red]{hyperref}
\usepackage{mathrsfs}
\usepackage{soul}
\usepackage{wrapfig}
\usepackage{xy} 
\usepackage{xcolor}
\usepackage{amsmath,amssymb,amsfonts,amsthm}
\usepackage[ascii]{inputenc}
\usepackage{makecell}
\allowdisplaybreaks

\newcommand{\be}{\begin{equation}}
	\newcommand{\ee}{\end{equation}}
\newcommand{\ba}{\begin{aligned}}
	\newcommand{\ea}{\end{aligned}}
\newcommand{\bea}{\begin{eqnarray}}
	\newcommand{\eea}{\end{eqnarray}}	
	
\newcommand{\beal}{\begin{align}}
\newcommand{\eal}{\end{align}}

\graphicspath{{images/}{../Figures}}

\begin{document}
%%%%%%%%%%%%%%%%%%%%%%%%%%%

%%%%%% TITLE %%%%%%%%%%%%%%
\title[Gap probability and full counting statistics in the $1d$OCP]{Gap probability and full counting statistics in the one dimensional one-component plasma}
%%%%%%%%%%%%%%%%%%%%%%%%%%%

\author{Ana Flack}
\address{LPTMS, CNRS, Univ.  Paris-Sud,  Universit\'e Paris-Saclay,  91405 Orsay,  France}
\author{Satya N. Majumdar}
\address{LPTMS, CNRS, Univ.  Paris-Sud,  Universit\'e Paris-Saclay,  91405 Orsay,  France}
\author{Gr\'egory Schehr}
\address{Sorbonne Universit\'e, Laboratoire de Physique Th\'eorique et Hautes Energies, CNRS UMR 7589, 4 Place Jussieu, 75252 Paris Cedex 05, France}

%%%%%%%%%%%%%%%%%%%%%%%%%%%

%%%%%%%% ABSTRACT %%%%%%%%%

\begin{abstract}
We consider the $1d$ one-component plasma (OCP) in thermal equilibrium, consisting of $N$ equally charged particles on a line, with pairwise
Coulomb repulsion and confined by an external harmonic potential. We study two observables: (i) the distribution of the gap between two consecutive particles in the bulk and 
(ii) the distribution of the number of particles $N_I$ in a fixed interval $I=[-L,+L]$ inside the bulk, the so-called full-counting-statistics (FCS). For both observables, we compute, for large $N$, the distribution of the typical as well as atypical large fluctuations. We show that the distribution of the typical fluctuations of the gap are described by the scaling form ${\cal P}_{\rm gap, bulk}(g,N) \sim N H_\alpha(g\,N)$, where $\alpha$ is the interaction coupling and the scaling function $H_\alpha(z)$ is computed explicitly. It has a faster than Gaussian tail for large $z$: $H_\alpha(z) \sim e^{-z^3/(96 \alpha)}$ as $z \to \infty$. Similarly, for the FCS, we show that the distribution of the typical fluctuations of $N_I$ is described by the scaling form ${\cal P}_{\rm FCS}(N_I,N) \sim 2\alpha \, U_\alpha[2 \alpha(N_I - \bar{N}_I)]$, where $\bar{N}_I = L\,N/(2 \alpha)$ is the average value of $N_I$ and the scaling function $U_\alpha(z)$ is obtained explicitly. For both observables, we show that the probability of large fluctuations are described by large deviations forms with respective rate functions that we compute explicitly. Our numerical Monte-Carlo simulations are in good agreement with our analytical predictions.    
\end{abstract}

%\usepackage[bb=boondox]{mathalfa}

%\title{Full counting statistics and gap probabilities in one dimensional one-component plasma}
%\author{}
%\begin{document}

\maketitle

\section{Introduction}
The Riesz gas is one of the most general and widely applicable models, ideal for the study of particles governed by long-range interactions
\cite{marcelriesz1938} -- for recent reviews with mathematical perspectives see \cite{LS2017,Lewine2022}. It is a system composed of $N$ particles that interact via pairwise interactions and are confined by an external harmonic potential.  The pairwise repulsive interactions vary with the distance as a power law and therefore the energy of the gas is given by
\begin{equation}
    E(\{x_i\}) = A\sum_{i=1}^{N}y_i^2 + \text{sgn}(k)B\sum_{i\neq j}\frac{1}{|y_i-y_j|^k},\quad A,B>0. 
    \label{eq:in:Riesz}
\end{equation}
where $\{y_i\}$ with $i\in \{1, \ldots , N\}$ denote positions of particles.  The parameter $k$ determines the strength of the interaction and should be greater than $-2$. Note that
the factor ${\rm sgn}(k)$ ensures that the pairwise interaction is always repulsive for $k>-2$. The lower bound $k>-2$ is such that the quadratic potential can confine the particles. 
For $k<-2$, the particles fly-off to $\pm\infty$.

Various special integer values of $k$ have been studied before. For example, by setting $k\rightarrow 0^+$, which results in a pairwise repulsion that varies as the logarithm of the distance,
the Dyson's log-gas is recovered \cite{Dyson1962, Mehtabook,Forrester}. Its special property is that the positions of the particles can be mapped to the eigenvalues of the Gaussian random matrices. The connection is made by identifying the Boltzman weight of the gas with the joint distribution of eigenvalues of an $N\times N$ random matrix belonging to one of the Gaussian ensembles of random matrices. Another known model that is part of the Riesz gas family is the classical Calogero-Moser model. Its energy is given by Eq.~\eqref{eq:in:Riesz} with $k=2$ and it is exactly solvable \cite{Calogero75, Moser76,AKD19}.\par 
In this article we study the case $k=-1$ which corresponds to one dimensional one component plasma ($1d$OCP) also known as the jellium model. The charges confined to a one-dimensional line interact by the Coulomb interaction, which in one dimension becomes linear pairwise repulsion between charges. The energy for $k=-1$ in Eq. (\ref{eq:in:Riesz}) reads
\begin{equation}
     E[\{y_i\}] = A\sum_{i=1}^{N}y_i^2 - B\sum_{i\neq j}|y_i-y_j| \;.
    \label{eq:in:energy_1dOCP}
\end{equation}
The first term, due to the confining harmonic potential, tends to push the particles towards the origin. It can be understood as an effective potential created by an uniform background of oppositely charged particles that guarantee charge neutrality. The second term, represented by 
a linear pairwise interaction, has an opposite effect, namely it pushes the charges away from each other. As a consequence of this competition between these two terms, 
the charges, on an average, settle down over a finite region of space in the limit of large $N$~\cite{Lenard, Prager, Baxter1963, Choquard1981}. 
In order that these two terms are of the same order, and the support of the density is $O(1)$ in the large $N$ limit, the coupling constants have to be rescaled with $N$. To estimate the rescaling factor, let us evaluate the two terms separately. We start with the first term and rescale the positions $y_i = c_N x_i$ where
$x_i = O(1)$ for large $N$. Then the first term scales as
\begin{equation}
    E_1 = \sum_{i=1}^{N}y_i^2 =c_N^2 \sum_{i=1}^N x_i^2 \sim c_N^2 N \;.
    \label{eq:int:T1}
\end{equation}
Similarly, the second term can be estimated as  
\begin{equation}
    E_2 = \sum_{i\neq j}^{N}|y_i -y_j| = c_N  \sum_{i\neq j}^{N}|x_i -x_j| \sim c_N N^2 \;,
    \label{eq:int:T2}
\end{equation}
where we used the fact that there are $N(N-1) \sim N^2$ terms in the double sum. Equating these two terms $E_1 \sim  E_2$, gives $c_N = O(N)$.  
Hence, in these scaled coordinates $x_i = O(1)$, the energy can be re-written as
\begin{equation} \label{betaE}
    \beta E[\{x_i\}] = \frac{N^2}{2}\sum_{i=1}^{N}x_i^2 - N\alpha \sum_{i\neq j}|x_i-x_j| \;,
\end{equation}
where $\alpha>0$ is the strength of the pairwise repulsion and we have, for convenience, expressed the energy in units of the inverse
temperature $\beta = 1/(k_B\,T)$. We assume that the gas is at thermal equilibrium at inverse temperature $\beta$, so that the probability 
of a configuration $\{x_1, x_2, \cdots, x_N\}$ of the positions of the charges is given by the Boltzmann-Gibbs weight 
\begin{equation} \label{GB}
  \mathcal{P}(\{x_i\}) =   \frac{1}{Z_N} e^{-\beta E[\{x_i\}]} \;,
\end{equation}
with $\beta E[\{x_i\}]$ is given in Eq.~\eqref{betaE}. Here $Z_N$ is the partition function of the gas, normalising the probability distribution, 
\bea \label{ZN_1}
Z_N = \int_{-\infty}^\infty dx_1  \int_{-\infty}^\infty dx_2 \cdots  \int_{-\infty}^\infty dx_N \, e^{-\beta E[\{x_i\}]} \;. 
\eea

The statistical mechanics of this simple model of a many-body interacting system has been of interest for many years. Due to the competing terms many intriguing effects emerge, but at the same time the system is simple enough that allows to extract analytical results for several observables~\cite{Lenard, Prager, Baxter1963, Choquard1981, Dean, Tellez, Dhar2017, Dhar2018, Flack2021}. This includes, for example, the distribution of the position of the rightmost particle \cite{Dhar2017, Dhar2018}, the number of particles with positive positions, the distribution of the gap between the rightmost and the next-to-rightmost particle~\cite{Dhar2018}, and also the so-called truncated linear statistics describing the distribution of the center of mass of $M$ rightmost particles~\cite{Flack2021}. Perhaps, the most natural observable for such a gas is the average macroscopic density of particles, defined as
\begin{equation}
    \rho_N(x) = \left \langle \frac{1}{N}\sum_{i=1}^{N}\delta(x-x_i) \right \rangle \;,
    \label{eq:int:density}
\end{equation}
where $\langle \ldots \rangle$ denotes an average over the Boltzmann-Gibbs measure in (\ref{GB}). In the large $N$ limit, the average density is known to converge to a $N$-independent uniform distribution over a finite support $[-2\alpha, 2 \alpha]$, namely~\cite{Lenard, Prager, Baxter1963, Choquard1981,Dhar2017} 
\begin{equation}  
   \lim_{N \to \infty} \rho_N(x) = \bar{\rho}(x) = \frac{1}{4\alpha},\quad -2\alpha \leq x \leq 2\alpha \;.
    \label{eq:in:one}
\end{equation}
In fact it can be shown that the flat average density corresponds to the minimum energy configuration in the large $N$ limit of the OCP gas in any dimension~\cite{Chafai,CFLV2017}. Coulomb interactions in different dimensions are given by the solutions of the Laplace equation. In one dimension this gives the linear repulsion, in two-dimensional space the solution is the logarithmic interaction and in higher dimensions the solution is $1/|x-y|^{d-2}$, where $d$ is the dimension of the space. For the OCP gas in $d$ dimensions, some observables, beyond the density, have been  
studied using Coulomb gas techniques. In particular, in $d=2$, where the Riesz gas with the logarithmic repulsion (corresponding to the $k\to 0^+$ limit in Eq.~(\ref{eq:in:Riesz})) is isomorphic to 
the Ginibre ensemble of random matrices \cite{Mehtabook,Forrester}, several recent results were derived analytically~\cite{ATW2014,ASZ2014,CFLV2017,LGMS18,BG18, Lacroix, Mex}. However, for the Riesz gas with index $k \neq d-2$ in Eq. (\ref{eq:in:Riesz}), even the computation of the averaged density is highly nontrivial for arbitrary $d$, including in $d=1$. The result for the density in arbitrary $d$ 
was known for a while for $k>1$ \cite{Lewine2022,GS72,MP72,Hardin18} and very recently it has been computed explicitly for all $k> -2$ in $d=1$~\cite{Agarwal2019}. Furthermore, these results were also extended to a class of systems with finite-range interactions \cite{KKK20}.

This average flat density in the large $N$ limit in $1d$, where the charges are equispaced, also corresponds to the configuration of charges that has the lowest energy.  
To see this, we order the positions of the charges $x_1<x_2< \cdots <x_N$, which conveniently gets rid of the absolute value $|x_i-x_j|$ in the energy. In terms
of these ordered positions, the energy can be rewritten as \cite{Baxter1963,Dhar2017,Dhar2018,Flack2021}
\begin{equation}
    \beta E[\{x_i\}] = N^3\left[\frac{1}{2N}\sum_{i=1}^{N}\left(x_i-\frac{2\alpha}{N}(2i-N-1)\right)^2\right] -C_N(\alpha) \;,
    \label{eq:in:energy_simpler}
\end{equation}
where
\begin{eqnarray}\label{defCN}
C_N(\alpha) = 2 \alpha^2 \sum_{i=1}^N (2i-N-1)^2 =  \frac{2\alpha^2}{3}N^3- \frac{2}{3}\,\alpha^2\,N \;.
\end{eqnarray}
The configuration with the lowest energy is the ``crystal configuration'' with equidistantly spaced charges at the positions
\begin{equation}
x_i=x_i^* = \frac{2\alpha}{N}(2i-N-1) \;.
\label{eq:in:positions}
\end{equation}
The partition function in (\ref{ZN_1}) is dominated, in the large $N$ limit, by the lowest energy configuration (\ref{eq:in:energy_simpler}) 
\bea \label{ZN_2}
Z_N \sim e^{-\beta E[\{x_i^* \}]} \sim e^{\frac{2}{3}\,\alpha^2\,N^3 + o(N^3)}  \;.
\eea
For later use, let us note that the partition function for finite $N$ can be written as a multiple $N$-fold
integral in terms of the ordered coordinates  
\begin{equation}
    Z_N = N!\int_{-\infty}^{\infty}dx_1\cdots \int_{-\infty}^{\infty}dx_Ne^{-\beta E[\{x_i\}]}\prod_{j=2}^N\theta(x_j-x_{j-1})\;,
    \label{eq:int:Zn}
\end{equation} 
where $\theta(z)$ is the Heaviside theta function that enforces the ordering of the positions of the particles.

The purpose of this paper is to present analytical results for two new observables in this $1d$OCP system: (i) the distribution of the gap between the positions of two consecutive particles deep inside the bulk of the density, far from the two edges $\pm 2 \alpha$ of the support of the average density, (ii) the full-counting-statistics (FCS), i.e., the 
distribution of the number of particles in an interval $[-L, + L]$, also in the bulk of the density. Both these observables are studied in the large $N$ limit where we compute both the typical behavior
as well as atypical ``large fluctuations behavior'', as explained in detail below. We also perform numerical simulations to verify our analytical predictions and find an excellent agreement.  

The rest of the paper is organized as follows. In Section II, we summarize our main results for the two observables (i) and (ii) discussed above.  
In Section III, we study the distribution of the bulk gap. Since it is easier to obtain the large deviation behavior of the bulk gap distribution using Coulomb gas techniques, we discuss first this large deviation regime in Section 3.1. In Section 3.2, we derive the typical behavior of the bulk gap distribution and also compare our analytical predictions to numerical simulations. In Section 4, we study the FCS, first in the large deviation regime in Section 4.1, followed by the typical regime in Section 4.2. Finally, we conclude in Section 5.

\section{Summary of the main results}\label{Sec:summary}

In this section, we summarize our main results on the observables (i) and (ii) discussed above.

\subsection{Distribution of the bulk gap}

The first observable concerns the gap $g_i = x_{i+1}-x_i$
between the positions of the $i$-th and the $(i+1)-th$ ordered particles. The case $i=1$ or symmetrically $i=N-1$ corresponds respectively to the first gap (close to the left edge 
of the support $-2 \alpha$ of the density) or symmetrically the gap closest to the right edge $+2\alpha$ and the distribution of these ``edge gaps'' have been studied before in the large $N$ limit in \cite{Dhar2018}. Here, we study instead the statistics of the gap in the bulk, i.e., far away from the two edges $\pm 2 \alpha$. To appreciate the difference between our results for the bulk gap from those at the edges, we first recall below the known results for the edge gap and then summarize our results for the bulk gap.

\vspace*{0.5cm}
\noindent {\it Known results for the statistics of the edge gap:} It was found that the typical fluctuations of the edge gap $g_1 = g_{N-1} \sim O(1/N)$ for large $N$, while the atypical large fluctuations are of order $O(1)$. These two behaviors are summarised as follows \cite{Dhar2018}
\bea \label{sum_g_edge}
 \mathcal{P}_{{\rm gap},{\rm edge}}(g, N) \approx
\begin{cases}
& N\,h_{\alpha}(g\,N) \;, \; g = O(1/N) \quad \quad ({\rm typical}) \;,\\
& \\
& e^{-N^2\psi_{\rm edge}(g)} \;, \; \quad g = O(1) \quad \quad \quad \;\; ({\rm atypical}) \;,
\end{cases}
\eea 
where the scaling function $h_\alpha(z)$ is given exactly by
\begin{equation}
    h_{\alpha}(z) = \theta (z)A(\alpha)\int_{-\infty}^{\infty}dy(y+z-4\alpha)e^{-(y+z-4\alpha)^2/2}F_{\alpha}(y) \;.
     \label{eq:in:gap:ed2}
 \end{equation}
Here $A(\alpha)$ is a constant that depends on the interaction strength $\alpha$ and $F_{\alpha}(x)$ is a function that appears rather regularly in 
 the computation of several observables in the $1d$OCP~\cite{Baxter1963,Dhar2017,Dhar2018}. Indeed, $F_\alpha(x)$ is the  
 unique solution of the non-local differential equation \cite{Baxter1963,Dhar2017,Dhar2018} 
  \begin{equation}
    \frac{d F_{\alpha}(x)}{dx}=A(\alpha)F_{\alpha}(x+4\alpha)e^{-\frac{x^2}{2}} \;,
    \label{eq:intro:eigenvalue_eq}
\end{equation}
with the boundary conditions $F_\alpha(x \to \infty) \to 1$ and $F_{\alpha}(x \to -\infty) \to 0$. This equation can be thought of as an eigenvalue equation, with 
$A(\alpha)$ as the unique eigenvalue for which there exists a solution that satisfies both boundary conditions. The asymptotic behaviors of the function $F_\alpha(x)$
are given by
\begin{eqnarray} 
1 - F_{\alpha}(x) &\sim& e^{-x^2/2 + o(x^2)} \quad , \quad \quad \quad x \to + \infty\;, \nonumber \\
\quad F_{\alpha}(x) &\sim& e^{-|x|^3/(24 \alpha)+ o(x^3)} \quad, \,\quad x \to - \infty \;. \label{asympt_Fa}
\end{eqnarray}
The full explicit expression for $F_{\alpha}(x)$ is not known, however it is possible to determine the solutions of Eq.~\eqref{eq:intro:eigenvalue_eq} by using numerical methods.
 From the first line of Eq.~(\ref{sum_g_edge}), it follows that the typical size of the fluctuations of the edge gap is of order $O(1/N)$ and the scaling function $h_\alpha(z)$ has an asymptotic Gaussian tail 
\bea \label{tail_h}
h_\alpha(z) \sim e^{-\frac{z^2}{2}} \;.
\eea
Note this behavior is independent of the interaction strength $\alpha$ as it corresponds to the cost in the harmonic potential energy in pulling the rightmost particle at a distance $g$ from the next-to-rightmost one. The probability of the atypical fluctuations of the gap, for $g = O(1)$, is described by the second line of Eq. (\ref{sum_g_edge}) where the rate function $\psi_{\rm edge}(g)$ is given by
 \bea \label{psi_gap}
 \psi_{\rm edge}(g) = \frac{g^2}{2} \;, \; g \geq 0 \;.
 \eea

\vspace*{0.5cm}
\noindent {\it New results for the statistics of the bulk gap:} A natural question is thus: how does the gap distribution behave as we go away from the two edges, towards the bulk of the gas? In this paper, we show that, while the typical size of the bulk gap still scales as $\sim O(1/N)$ and the atypical large bulk gap $\sim O(1)$ (as in the case of the edge gaps), the corresponding distributions of the gap in the bulk 
are entirely different from those at the edges. More precisely, we show that the analog of Eq. (\ref{sum_g_edge}) now reads
\begin{equation}
    \mathcal{P}_{{\rm gap}, {\rm bulk}}(g, N) \sim
    \begin{cases}
    &NH_{\alpha}(g \, N), \quad g\sim O(1/N),\\
    &\\
    &e^{-N^3\psi_{\rm bulk}(g)}, \quad g\sim O(1) \;,
    \end{cases}
    \label{eq:intro:gap_distribution}
\end{equation}
where the scaling function $H_{\alpha}(z)$ is given by 
\begin{equation}
    {
    H_{\alpha}(z)= \theta(z) B\, A^2(\alpha) \int_{-\infty}^{\infty} dy F_{\alpha}(y+4\alpha) F_{\alpha}(8 \alpha -y-z) e^{-y^2/2-(y+z-4\alpha)^2/2}} \;,
    \label{eq:intro:gap_scaling_f}
\end{equation}
where the function $F_{\alpha}(x)$ and  the eigenvalue $A(\alpha)$ are determined from Eq. (\ref{eq:intro:eigenvalue_eq}) and $B$ is a normalization constant such that 
\begin{equation}
 \int_0^{\infty}\mathcal{P}_{{\rm gap}, {\rm bulk}}(g, N) \, dg = \int_0^\infty   H_{\alpha}(z) \, dz = 1 \;.
    \label{eq:intro:gap_normalization}
\end{equation}
The scaling function $H_\alpha(z)$ behaves, for large $z$, as
\bea \label{asympt_H}
H_{\alpha}(z) \sim e^{-\frac{z^3}{96\alpha} + o(z^3)} \quad, \quad z \to + \infty \;.
\eea
Comparing to the edge gap asymptotics in Eq. (\ref{tail_h}), we notice two properties: first, the tail of the scaling function $H_\alpha(z)$ is highly non-Gaussian in Eq. (\ref{asympt_H}), as opposed to the Gaussian tail of the edge gap. Secondly, the tail of $H_\alpha(z)$ in Eq. (\ref{asympt_H}) depends explicitly on the interaction strength $\alpha$. Thus this tail emerges from the strong interaction between the bulk particles and hence the physics of the gap in the bulk differs drastically from that at the edges. 

The second line of Eq. (\ref{eq:intro:gap_distribution}) describes the atypical fluctuations of order $O(1)$ of the bulk gap where the rate function $\psi_{\rm bulk}(g)$ is given explicitly by
\begin{equation}
{\psi_{\rm bulk}(g) = \frac{g^3}{96\alpha}} \quad, \quad g \geq 0 \;.
    \label{eq:intro:gap_ldf}
\end{equation}
Substituting the asymptotic behavior of the scaling function $H_\alpha(z)$ for large $z$ from Eq. (\ref{asympt_H}) into the first line of Eq. (\ref{eq:intro:gap_distribution}), one finds that $P_{{\rm gap}, {\rm bulk}}(g,N) \sim e^{-(g^3 N^3)/(96 \alpha)}$, which matches smoothly with the large deviation behavior displayed in the second line of Eq. (\ref{eq:intro:gap_distribution}).  Thus, at the level of large deviations also, we see that the behavior of the bulk gap differs from that of the edge: the large deviation function of the edge gap is quadratic in $g$ in Eq. (\ref{psi_gap}) while it is cubic in $g$ in Eq. (\ref{eq:intro:gap_ldf}) for the bulk~gap.  

\vspace*{0.5cm}

\subsection{Full counting statistics}

The second observable that we study in this paper is the FCS, i.e., the distribution of the number of particles $N_I$ in the interval $[-L,+L]$. The FCS has been studied in many different systems, such as in the study of quantum shot noise, transport and quantum dots \cite{Levitov, Levitov2, Groth, Gustavsson} as well as spin chains \cite{Ivanov, Eisler, Stephan, Groha, Gamayun}. Furthermore, it has been well studied for the eigenvalues of certain ensembles of random matrices \cite{Forrester,CL1995,FS1995,Marino2014,Marino2016}. These results are particularly interesting since the eigenvalues of unitary Gaussian ensemble can be mapped to non-interacting fermions in a harmonic trap \cite{Marino2014, Marino2016, Dean2018}. The variance of FCS is an interesting observable since it characterizes quantum fluctuations in the ground state. It has been computed for different systems of interacting and non-interacting trapped fermions \cite{Lacroix, Smith2020A, Smith2021, Smith2021A,Smith2021B,Gouraud22}. In this paper, we show that the FCS can be computed analytically for large $N$, in the $1d$OCP model. 
For each equilibrium realisation of the $1d$OCP, the random variable $N_I$ can be represented~as 
 \begin{equation}
    N_I = \sum_{l=1}^{N}\mathbb{I}_{I}(x_l),
    \label{eq:int:NI}
\end{equation}
where $\mathbb{I}_{I}(x_l)$ is an indicator function which takes the value $1$ if the $i$-th particle is inside the interval $I$ and zero otherwise.  
Since the equilibrium density of charges is uniform on the support $[-2\alpha, 2\alpha]$, the average number of particles in an interval of length $2L$ is
\begin{equation}
    \bar{N}_I = \frac{2L}{4\alpha}N=\frac{L}{2\alpha}N \;,
    \label{eq:intro:average_FCS}
\end{equation}
where we have assumed that $L < 2\alpha$. We study here the probability distribution of the typical and large fluctuations of $N_I$ around its mean $\bar{N}_I$. We show that the typical size of the fluctuations 
$N_I - \bar{N}_I$ is of order $O(1)$ for large $N$, while the large fluctuations are of order $O(N)$. These two behaviors are encoded in the following scaling forms
\begin{eqnarray}
    \label{eq:int:FCS1}
\hspace*{-1cm}    \mathcal{P}_{FCS}(N_I, N) \sim 
    \begin{cases}
    &2\alpha U_{\alpha}[2\alpha\,({N}_I-\bar{N}_I)], \quad |{N}_I-\bar{N}_I| = O(1) \;, \; ({\rm typical})\\
    & \\
    &e^{-N^3\Phi(N_I/N)} \quad, \quad\quad\quad\;\; |{N}_I-\bar{N}_I| = O(N) \;, \;  ({\rm atypical}) \;,
    \end{cases}
\end{eqnarray}
where the scaling function $U_\alpha(z)$ is given by
\bea
U_{\alpha}(z) = \frac{F_{\alpha}^2(-z+2\alpha) F_{\alpha}^2(z+2\alpha)}{\int_{-\infty}^\infty F_{\alpha}^2(-z+2\alpha) F_{\alpha}^2(z+2\alpha) \, dz}   \;,
    \label{eq:intro:FCS_typical}
\eea
where the same function $F_{\alpha}(x)$ defined in Eq. (\ref{eq:intro:eigenvalue_eq}) appears again. The function $U_\alpha(z)$ is symmetric in $z$ and it behaves, for large $|z|$, as
\bea \label{largeK}
U_\alpha(z) \sim e^{-\frac{|z|^3}{12 \alpha}+ o(|z|^3)} \;.
\eea
The rate function $\Phi(z)$ in the second line of Eq. (\ref{eq:int:FCS1}) is given explicitly by
\begin{equation}
 {
    \Phi(z) = \frac{2}{3}\alpha^2\, \left|z-\frac{L}{2\alpha}\right|^3} \;.
    \label{eq:intro:LDF_counting_stat}
\end{equation}
Substituting the asymptotic large $|z|$ behavior of $U_\alpha(z)$ from Eq. (\ref{largeK}) into the first line of Eq. (\ref{eq:int:FCS1}), one obtains, for large fluctuations
compared to the typical one, $\mathcal{P}_{FCS}(N_I, N) \sim e^{- \frac{2}{3}\alpha^2 |N_I - \bar{N}_I|^3}$ which matches smoothly with the large deviation behavior described in the second line of Eq. (\ref{eq:int:FCS1}) and (\ref{eq:intro:LDF_counting_stat}). Finally, we note that, quite remarkably, the rate function $\Phi(z)$ in Eq. (\ref{eq:intro:LDF_counting_stat}) is non-analytic at its minimum $z = L/(2\alpha)$.

%\newpage

 \par 
\begin{figure}
        \centering
        \includegraphics[width=0.7\linewidth]{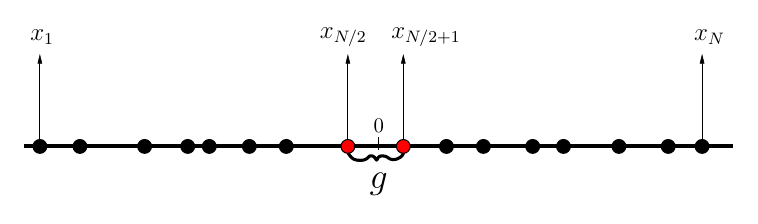}
        \caption{A symbolic example of ordered configuration of 1dOCP. The particles are labeled from left to right so that $x_1<x_2< \cdots <x_N$. The bulk gap $g$ is given by the difference of positions of two middle particles $g = x_{N/2+1}-x_{N/2}$.}
        \label{fig:gap:typical:picture}
\end{figure}

\section{Distribution of the bulk gap }

In this section, we compute the distribution of the gap $g_i = x_{i+1}-x_i$ between the $i$-th and $(i+1)$-th particle in the bulk. Using the equilibrium positions of the charges in the large $N$ limit in 
Eq. (\ref{eq:in:positions}), it follows that the average gap in the bulk is 
\bea \label{av_gap}
\langle g_i \rangle \simeq \frac{4\alpha}{N} \;,
\eea
which, clearly, is independent of the index $i$. However the gap $g_i$ is a random variable and fluctuates from sample to sample around its mean value $4\alpha/N$. Our goal is to compute the distribution of these fluctuations, both typical and large, for large $N$. If the index $i$ of the particle is $i=cN$ where $c=O(1)$, then the particle with position $x_i$ is in the bulk of the Coulomb gas. In contrast, if $i = O(1)$ or $i \approx N$, the particles are close to the two edges. We expect that the statistics of $g_i$ is independent of $i$ as long as the gap $g_i$ is in the bulk, while it will be different if the $g_i$ is close to the two edges. As already discussed in Section \ref{Sec:summary}, the edge gap distribution was computed in Ref. \cite{Dhar2018}. Here our goal is to compute the distribution of $g_i$ in the bulk. Clearly, the statistics of $g_i$ in the bulk should not depend on the index $i$, as long as $1 \ll i \ll N$. It is then convenient to set $i = N/2$ and hence consider the mid-gap
\begin{equation}
g = x_{N/2+1}-x_{N/2} \;. \label{eq:gap:g}
\end{equation}
We will see that the scale of typical fluctuations is of order $O(1/N)$ around its mean value $\langle g \rangle = 4\alpha/N$, while the large fluctuations are
of order $O(1)$ around the mean. Below, in Subsection 3.1, 
we first discuss the large deviation regime, where the calculation is simpler using the Coulomb-gas technique. Then, in Subsection 3.2, we discuss the typical fluctuations.

    \subsection{Large fluctuations of the gap}\label{sec:bulkgap}
    \begin{figure}
        \centering
        \includegraphics[width=\linewidth]{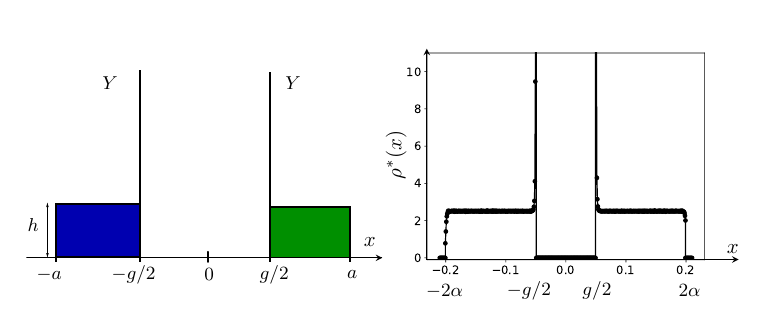}
        \caption{Density of the $1d$OCP with the additional constraint that the positions of the two central particles are $x_{N/2} = -g/2$ and $x_{N/2+1} = +g/2$ leading to a mid-gap $g$ in the bulk [see Eq. (\ref{eq:gap:g})]. On the left panel we provide a schematic depiction of the ansatz for the saddle point density in Eq. (\ref{eq:gap:large:ansatz}), which consists of two symmetric flat blocks over $[-a,-g/2]$ and $[+g/2,+a]$ with height $h$ in each block and, in addition, two delta peaks of amplitude $Y$ located at $\pm g/2$. As argued in the text [see Eqs. (\ref{Stot_sp}) and (\ref{Ystar})], the saddle point density has parameter values $a=a^* = 2\alpha$, $h = h^*=1/(4 \alpha)$ and $Y =Y^*= g/(8\alpha)$. 
On the right panel we show the result of Monte-Carlo simulations with $N=1000$ particles and gap of size $g=0.1$, with coupling parameter $\alpha=0.1$. This would correspond to $a^*=0.2$, $h^* = 2.5$ and $Y^* = 0.125$. Simulation confirms the saddle point ansatz depicted schematically on the left panel.}
        \label{fig:gap:large:picture}
    \end{figure}
    
        The starting point for analyzing the gap statistics is the exact expression of the distribution given by the $N$-fold integral 
        \begin{eqnarray}\label{eq:gap:large:distribution}
        \hspace*{-2.5cm}    \mathcal{P}_{{\rm gap}, {\rm bulk}}(g, N) &=&
            \frac{\int_{-\infty}^{\infty}dx_1...\int_{-\infty}^{\infty}dx_N e^{- N^3\, S[\{x_i\}]}\delta(g-(x_{N/2+1}-x_{N/2}))\prod_{i=2}^N\theta(x_{j}-x_{j-1})}{\int_{-\infty}^{\infty}dx_1...\int_{-\infty}^{\infty}dx_N e^{- N^3\, S[\{x_i\}]}\prod_{i=2}^N\theta(x_{j}-x_{j-1})} \nonumber \\
            &=& \frac{{\cal N}_N}{{\cal D}_N} \;,
        \end{eqnarray}
where the $x_i$'s are the ordered positions of the charges and the action $S[\{x_i\}] = \beta E[\{x_i\}]/N^3$ with $\beta E[\{x_i\}]/N^3$ in Eq. (\ref{eq:in:energy_simpler}). This action then reads explicitly, to leading order for large $N$, 
          \begin{equation}
        S[\{x_i\}]= \frac{1}{2N}\sum_{i=1}^{N}\left(x_i-\frac{2\alpha}{N}(2i-N-1)\right)^2  -  \frac{2 \alpha^2}{3} \;.
        \label{eq:gap:large:action}
        \end{equation}
Thus the exact gap distribution in (\ref{eq:gap:large:distribution}) is a ratio of two partition functions: the numerator ${\cal N}_N$ corresponds to the partition function of the jellium model under the additional constraint ensuring that the distance between two charges at the center of the gas is equal to $g$, while the denominator is the partition function of the unconstrained model.        
        
Our goal is to evaluate the multiple integrals in the numerator and in the denominator of Eq. (\ref{eq:gap:large:distribution}) by a saddle point method for large $N$. Consider the denominator first. In this case, the saddle point occurs at the $x_i = x_i^* = 2\alpha/N(2i-N-1)$ as given in Eq. (\ref{eq:in:positions}). Thus, in the large $N$ limit, the saddle point density is given by the flat distribution $\bar{\rho}(x) = 1/(4\alpha)$ (for $-2\alpha \leq x \leq 2 \alpha$). This gives the denominator, to leading order for large $N$ 
\begin{eqnarray} \label{DN}
{\cal D}_N \approx e^{\frac{2\alpha^2}{3}N^3} \;.
\end{eqnarray}        
We now want to evaluate the numerator ${\cal N}_N$ in Eq. (\ref{eq:gap:large:distribution}) also by a saddle-point method for large $N$. 
Due to the presence of the additional gap constraint, we expect the saddle point density to break into two symmetrically  
disjoint parts (with each containing half the charges), with a separation $g$ between them. 
To guess the saddle point density more precisely, we first performed a 
Monte-Carlo simulation. These simulations showed that in the presence of the gap $g>0$: (i) the solution for the saddle-point density consists of charges distributed uniformly over two symmetric blocks, blue on the left and green on the right, as seen in the left panel of Fig. \ref{fig:gap:large:picture}, over a support $[-a,-g/2]$ (left) and $[g/2,a]$ (right), (ii) the height of the flat density in each block is some value $h>0$ and (iii) the displaced charges due to the creation of the gap accumulate at the two inner edges as $\delta$-peaks with equal intensity $Y$
(see the right panel of Fig.~\ref{fig:gap:large:picture}). This leads us to make the following ansatz for the saddle-point density 
        \begin{eqnarray}
            \rho^*(x)=\begin{cases}
                &h+Y\,\delta\left(x + \frac{g}{2}\right)\, \quad, \quad -a \leq x \leq -\frac{g}{2},\\
                & \\
                &h+Y\,\delta \left(x - \frac{g}{2}\right) \,\quad, \quad\quad \frac{g}{2}\leq x \leq a \;.
                 \end{cases}
            \label{eq:gap:large:ansatz}
        \end{eqnarray}
In addition, the total number of charges in the left block is $N/2$ (and similarly on the right). Using the uniform density $h$ in each block, this gives the condition
\begin{eqnarray} \label{norm}
Y + h \left(a - \frac{g}{2} \right) = \frac{1}{2} \;.
\end{eqnarray} 
We now evaluate the action in Eq. (\ref{eq:gap:large:action}) using this saddle-point ansatz for the charge densities. First, we consider the contribution from the uniform background on the left block (i.e., without the delta peak at $-g/2$). To evaluate the action, we note that the uniform density of height $h$ provides a relation between the position $x_i$ of the $i$-th particle in the left block
\begin{eqnarray} \label{rel_xi}
-a + \frac{1}{hN}i = x_i  \;,
\end{eqnarray}
where we used the fact that the inter-particle distance in the left block is $1/(hN)$. Substituting the expression of $x_i$ from Eq. (\ref{rel_xi}) in the action (\ref{eq:gap:large:action}), 
we get (without including the constant factor $-2 \alpha^2/3$)
\begin{eqnarray}\label{Sleft}
S_{\rm left, {\rm bulk}} = \frac{1}{2N} \sum_{i=1}^{N/2-Y N} \left(\frac{i}{hN}-a - \frac{2\alpha}{N}(2i-N-1) \right)^2  \;,
\end{eqnarray}
where the upper limit of the summation $N/2-YN$ comes from the fact that there are $YN$ charges in the delta peak and hence the number of particles in the uniform background in the left block is $N/2-YN$. This sum can be evaluated in the large $N$ limit by replacing it by an integral over $z=i/N$. This gives
\begin{eqnarray}\label{Sleft_int}
S_{\rm left, {\rm bulk}}  \simeq \frac{1}{2}\,\int_0^{1/2-Y} \left( \left(\frac{1}{h}-4\alpha \right)\,z + 2 \alpha -a \right)^2 \, dz \;.
\end{eqnarray}
We now consider the contribution to the action from the delta peak in the left block at $-g/2$. Setting $x_i = - g/2$ in Eq. (\ref{eq:gap:large:action}), we get
\begin{eqnarray} \label{Sleft_delta}
S_{\rm left, {\rm delta}}  &=&  \frac{1}{2N} \sum_{i = N/2 - YN}^{N/2} \left( -\frac{g}{2} -\frac{2\alpha}{N}(2i-N-1)\right)^2 \nonumber \\
&\simeq& \frac{1}{2} \int_{1/2-Y}^{1/2} \left(-\frac{g}{2} - 4 \alpha z + 2\alpha  \right)^2 \, dz \;.
\end{eqnarray}
The contribution from the right block is exactly the same as the left block, using the symmetry. Hence the total contribution to the saddle-point action $S^*$, now including the constant factor $-2 \alpha^2/3$ from Eq. (\ref{eq:gap:large:action}), is given by
\begin{eqnarray}\label{Stot}
&&\hspace*{-2.5cm}S^* = 2 (S_{\rm left, {\rm bulk}} + S_{\rm left, {\rm delta}}) - \frac{2\alpha^2}{3}  \\
&&\hspace*{-2.cm}= \int_0^{1/2-Y} \left( \left(\frac{1}{h}-4\alpha \right)\,z + 2 \alpha -a \right)^2 \, dz +  \int_{1/2-Y}^{1/2} \left(-\frac{g}{2} - 4 \alpha z + 2\alpha  \right)^2 \, dz - \frac{2\alpha^2}{3} \;. \nonumber
\end{eqnarray}
We can now replace $Y$ in the action in (\ref{Stot}) in terms of $h$ and $a$ using Eq. (\ref{norm}). This gives the saddle point action $S^*$ in terms of two independent parameters $a$ and $h$. Minimizing the action with respect to $a$ and $h$, i.e., by setting $\partial_a S^* = 0$ and $\partial_g S^* = 0$, gives the desired saddle-point parameters $a^*$ and $h^*$. We carried out this minimisation using Mathematica -- since the integrals in (\ref{Stot}) are a bit cumbersome. The final solution is however amazingly simple
\begin{eqnarray} \label{Stot_sp}
a^* = 2 \alpha \quad, \quad h^* = \frac{1}{4\alpha} \;.
\end{eqnarray}
This means that the intensity of the delta-peaks, using Eq. (\ref{norm}), is
\begin{eqnarray}\label{Ystar}
Y^* = \frac{g}{8\alpha} \;.
\end{eqnarray}
Thus the saddle-point density is rather simple: when a gap is introduced in the bulk, it does not affect the bulk densities on either side since they stay at their unconstrained height $h^* = 1/(4\alpha)$ and also the support edges $\pm 2 \alpha$ also remain the same as the unconstrained case. The charges that get displaced due to the presence of the gap accumulate as delta-peaks at the edges of the gap, i.e., at $\pm g/2$. These results in Eqs. (\ref{Stot_sp}) and (\ref{Ystar}) are confirmed by our simulations shown on the right panel of Fig. \ref{fig:gap:large:picture}.  
%In fact, this charge density profile for the saddle point turns out to be rather generic for Coulomb gases in arbitrary dimensions. A gap in one-dimension or a hole in higher dimensions do not affect the bulk charge density and the displaced charges get distributed over the surface of the hole \cite{Dhar2017,Dhar2018,ATW2014,CFLV2017,Flack2021}. 
    
We now evaluate the action at the saddle-point density. Substituting $h=1/(4 \alpha)$, $a = 2 \alpha$ and $Y = g/(8\alpha)$ in Eq. (\ref{Stot}), it is easy to see that the first integral representing the bulk disappears. The rest gives
\begin{eqnarray} \label{Sstar}
S^* = \frac{g^3}{96 \alpha} - \frac{2 \alpha^2}{3} \;,
\end{eqnarray}     
leading to 
\begin{eqnarray} \label{Nstar}
{\cal N}_N \sim e^{-N^3\,S^*} \sim e^{-N^3(\frac{g^3}{96} - \frac{2\alpha^2}{3}) } \;.
\end{eqnarray}
Using the result for the denominator in Eq. (\ref{DN}), and taking the ratio in Eq. (\ref{eq:gap:large:distribution}) gives
\begin{equation}
        {
            \mathcal{P}_{\rm gap, bulk}(g, N)\sim e^{-N^3\psi_{\rm gap}(g)}, \quad \psi_{\rm gap}(g) = \frac{g^3}{96\alpha} \;,}
            \label{eq:gap:large:ldf}
        \end{equation}
as announced in Eq. (\ref{eq:intro:gap_ldf}) in the introduction.

    \subsection{Typical fluctuations of the gap} \label{Sec:typgap}

Our starting point again is the exact expression of the bulk gap distribution in Eq. (\ref{eq:gap:large:distribution}), but here we are interested to compute the {\it typical} fluctuations of the gap, of size
${\cal O}(1/N)$. Note that when the gap is of order $O(1)$, as in the previous subsection, one could use the saddle-point method because, introducing a gap of order $O(1)$ changes the macroscopic saddle-point density, leading to different saddle points for the numerator ${\cal N}_N$ and the denominator ${\cal D}_N$ in Eq. (\ref{eq:gap:large:distribution}). However, if the gap is of order $O(1/N)$, the macroscopic saddle-point density in the numerator and the denominator in Eq. (\ref{eq:gap:large:distribution}) 
remains the same in the large $N$ limit. Consequently, from Eq. (\ref{eq:gap:large:ldf}), $\psi_{\rm gap}(g)$ vanishes and the large deviation form in (\ref{eq:gap:large:ldf}) does not describe the distribution of the gap when $g = O(1/N)$. In fact, it turns out, as shown below, that the gap distribution has a different scaling form, $\mathcal{P}_{\rm gap, bulk}(g, N)\sim N H_\alpha(g\,N)$. Clearly, the saddle-point method is not suitable to extract this scaling form. For this, one needs to evaluate the integral in the numerator ${\cal N}_N$ in a more precise manner that brings out this scaling form.

Here, it is more convenient to slightly re-write Eq. (\ref{eq:gap:large:distribution}) as
\begin{equation}\label{P_gapZ}
\hspace*{-.cm}\mathcal{P}_{\rm gap, bulk}(g, N) = \frac{N!}{Z_N}  \int_{-\infty}^{\infty}dx_1 \ldots \int_{-\infty}^{\infty}dx_N e^{- \beta E[\{x_i\}]}\delta(g-(x_{N/2+1}-x_{N/2}))\prod_{i=2}^N\theta(x_{j}-x_{j-1}) \;,
\end{equation}
where the scaled energy $\beta E[\{ x_i\}]$ is given in Eq. (\ref{eq:in:energy_simpler}) and $Z_N$ in Eq. (\ref{eq:int:Zn}). We first introduce a change of variables  
        \begin{equation}
            \epsilon_i = Nx_i - 2\alpha (2i-N-1) \;,
            \label{eq:gap:typ:epsilon}
        \end{equation}
in terms of which the scaled energy in Eq. (\ref{eq:in:energy_simpler}) reduces to $\beta E[\{x_i\}] = \frac{1}{2}\sum_{i=1}^{N}\epsilon_i^2$. Consequently, the gap distribution reads
        \begin{eqnarray}
 &&      \hspace*{-2.5cm}     \mathcal{P}_{\rm gap, bulk}(g, N) =
            N^{1-N}\frac{N!\,e^{C_N(\alpha)}}{Z_N}\int_{-\infty}^{\infty}d\epsilon_1 \ldots \int_{-\infty}^{\infty}d\epsilon_N e^{-\frac{1}{2}\epsilon_k^2}\delta(\epsilon_{N/2+1}-\epsilon_{N/2}-gN+4\alpha) \nonumber \\
            &&\times \prod_{j=2}^{N}\theta(\epsilon_j-\epsilon_{j-1}+4\alpha) \;,
            \label{eq:gap:typ:prob2}
        \end{eqnarray}
where $C_N(\alpha)$ is given in Eq. (\ref{defCN}). Thus, in terms of the variables $\epsilon_i$'s, the gas becomes short-ranged.    
   
In the next step we separate out the two integrals over $\epsilon_{N/2+1}$ and $\epsilon_{N/2}$. The rest can be divided into two parts: the left block corresponding to 
integrals over $\epsilon_i$'s with $i < N/2$ and the right block  corresponding to $\epsilon_i$'s with $i > N/2+1$.  
        \begin{align}\label{eq:gap:typ:A}
            \mathcal{P}_{\rm gap, bulk}(g, N) &= \frac{N^{1-N}N!\,e^{C_N(\alpha)}}{Z_N} \nonumber \\
           \hspace*{-1cm}\times &\int_{-\infty}^{\infty}d\epsilon_{N/2+1}\int_{-\infty}^{\infty}d\epsilon_{N/2}e^{-\frac{1}{2}(\epsilon_{N/2+1}^2+\epsilon_{N/2}^2)}\delta(\epsilon_{N/2+1}-\epsilon_{N/2}-gN+4\alpha)\theta(\epsilon_{N/2+1}-\epsilon_{N/2}+4\alpha) \nonumber \\
           \times  &\int_{-\infty}^{\infty}\prod_{k=1}^{N/2-1}d\epsilon_k e^{-\frac{1}{2}\sum_{k=1}^{N/2-1}\epsilon_k^2}\prod_{k=2}^{N/2}\theta(\epsilon_k-\epsilon_{k-1}+4\alpha) \\
            \times &\int_{-\infty}^{\infty}\prod_{N/2+2}^Nd\epsilon_ke^{-\frac{1}{2}\sum_{k=N/2+2}^{N}\epsilon_k^2}\prod_{k=N/2+2}^{N}\theta(\epsilon_k-\epsilon_{k-1}+4\alpha)\nonumber.
        \end{align}
 The integrals belonging to the left (of $i = N/2$) and to the right (of $i=N/2+1$) blocks can be conveniently expressed in terms of two functions that were already introduced in Ref.~\cite{Dhar2018} for calculating the typical fluctuations of the gap near the edge, namely
        \begin{align}
            D_{\alpha}(x, n) &= \int_{-\infty}^{x}d \epsilon_n\int_{-\infty}^{\epsilon_n+4\alpha}d \epsilon_{n-1}...\int_{-\infty}^{\epsilon_2+4\alpha}d \epsilon_1 e^{-\frac{1}{2}\sum_{k=1}^{n}\epsilon_k^2},\\
            E_{\alpha}(x, n) &= \int_{x}^{\infty}d \epsilon_1\int_{\epsilon_1-4\alpha}^{\infty}d \epsilon_2 ... \int_{\epsilon_{n-1}-4\alpha}^{\infty}d \epsilon_n e^{-\frac{1}{2}\sum_{k=1}^{n}\epsilon_k^2} \;.
            \label{eq:gap:typ:ED}
        \end{align}
The function $D_\alpha(x,n)$ can be physically interpreted as the partition function of the gas (up to constant prefactors) confined in the domain $(-\infty, x]$, while $E_\alpha(x,n)$ corresponds
to the case where the confining domain is $[x, +\infty)$. There is an obvious symmetry connecting these two functions, namely       
        \begin{equation}
            E_{\alpha}(-x, n) = D_{\alpha}(x, n),
            \label{eq:gap:typ:E_D}
        \end{equation}
        which can be checked by changing the variables $\epsilon_k \rightarrow -\epsilon_k$ in the definition of $E_{\alpha}(x, n)$ in Eq. (\ref{eq:gap:typ:ED}). In terms of these two functions, one can easily see that Eq.~(\ref{eq:gap:typ:A}) reads
        \begin{align}
            \mathcal{P}_{\rm gap, bulk}(g, N) = \frac{N^{1-N}N!\,e^{C_N(\alpha)}}{Z_N}
            \int_{-\infty}^{\infty}d\epsilon_{N/2+1}\int_{-\infty}^{\infty}d\epsilon_{N/2}e^{-\frac{1}{2}(\epsilon_{N/2+1}^2+\epsilon_{N/2}^2)} \nonumber\\ 
            \times \delta(\epsilon_{N/2+1}-\epsilon_{N/2}-gN+4\alpha)\theta(\epsilon_{N/2+1}-\epsilon_{N/2}+4\alpha) \\
            \times D_{\alpha}(\epsilon_{N/2}+4\alpha, N/2-1)
            \times E_{\alpha}(\epsilon_{N/2+1}-4\alpha, N/2-1) \;. \nonumber
            \label{eq:gap:typ:B}
        \end{align}
Using the symmetry in Eq. (\ref{eq:gap:typ:E_D}), and making a change of variable $\epsilon_{N/2+1} \to - \epsilon_{N/2+1}$, this can be written as       
           \begin{eqnarray}\label{eq:gap:typ:C}
            \mathcal{P}_{\rm gap, bulk}(g, N) = \frac{N^{1-N}N!\, e^{C_N(\alpha)}}{Z_N}
            \int_{-\infty}^{\infty}d\epsilon_{N/2+1}\int_{-\infty}^{\infty}d\epsilon_{N/2}e^{-\frac{1}{2}(\epsilon_{N/2+1}^2+\epsilon_{N/2}^2)}\nonumber\\ 
            \times \delta(-\epsilon_{N/2+1}-\epsilon_{N/2}-gN+4\alpha)\theta(-\epsilon_{N/2+1}-\epsilon_{N/2}+4\alpha) \\
            \times D_{\alpha}(\epsilon_{N/2}+4\alpha, N/2-1)
            D_{\alpha}(\epsilon_{N/2+1}+4\alpha, N/2-1) \;.\nonumber
        \end{eqnarray}
So far, the result for the distribution of the gap in Eq. (\ref{eq:gap:typ:C}) is exact for any $N$. We then consider the scaling limit $N \to \infty$, $g \to 0$ but keeping the product $g \, N$ fixed. This limit picks up the contributions of the gap of size $O(1/N)$. To study the limit $N \to \infty$ of the function $D_\alpha(x,N)$, we follow Ref. \cite{Dhar2017} and 
introduce the ratio $F_{\alpha}(x,M)$ 
        \begin{equation}
            F_{\alpha}(x, M) = \frac{D_{\alpha}(x, M)}{D_{\alpha}(\infty, M)}  \;.
                        \label{eq:gap:typ:F}
        \end{equation}
The denominator $D_{\alpha}(\infty, M)$, by definition in Eq. (\ref{eq:gap:typ:ED}), is simply proportional to the partition function $Z_M$ of the $1d$OCP gas with $M$ particles. More precisely, it is easy to see that
 \begin{equation}
            Z_M = \frac{M! e^{C_M(\alpha)}}{M^{M}}D_{\alpha}(\infty, M) \;.
            \label{eq:gap:typ:partition}
        \end{equation}
The ratio $F_{\alpha}(x, M)$ in Eq. (\ref{eq:gap:typ:F}) then represents the cumulative probability that the position of the rightmost particle is less than $x$. In particular, $F_{\alpha}(x \to +\infty, M) = 1$ and $F_{\alpha}(x \to -\infty, M) = 0$. In terms of this function $F_\alpha(x,M)$, the gap distribution in (\ref{eq:gap:typ:C}),           
        \begin{eqnarray}
            \mathcal{P}_{\rm gap, bulk}(g, N) = N\frac{D_{\alpha}^2(\infty, N/2-1)}{D_{\alpha}(\infty, N)}
            \int_{-\infty}^{\infty}d\epsilon_{N/2+1}\int_{-\infty}^{\infty}d\epsilon_{N/2}e^{-\frac{1}{2}(\epsilon_{N/2+1}^2+\epsilon_{N/2}^2)}\nonumber\\ 
            \times \delta(-\epsilon_{N/2+1}-\epsilon_{N/2}-gN+4\alpha)\theta(-\epsilon_{N/2+1}-\epsilon_{N/2}+4\alpha) \nonumber \\
            \times F_{\alpha}(\epsilon_{N/2}+4\alpha, N/2-1)
            F_{\alpha}(\epsilon_{N/2+1}+4\alpha, N/2-1) \;,
            \label{eq:gap:typ:D}
        \end{eqnarray}
where we have used the relation (\ref{eq:gap:typ:partition}).    

We now consider the large $N$ limit in Eq. (\ref{eq:gap:typ:D}). First, we consider the large $M$ behavior of the factor $D_\alpha(\infty, M)$ that appears in the prefactor in (\ref{eq:gap:typ:D}). 
Since $D_\alpha(\infty, M)$ is the proportional to the partition function of a short-ranged gas, one expects that its free-energy $-\ln{D_{\alpha}(x, M)} \propto M$ is extensive in the number of 
particles $M$ for large $M$. This indicates that, to leading order for large $M$, 
 \begin{equation}
            D_{\alpha}(\infty, M) \sim B\, [A(\alpha)]^{-M} \;,
            \label{eq:gap:typ:R}
        \end{equation}       
where we assume that $B$ (to be verified a posteriori) is an $M$-independent constant. Note that $-\ln A(\alpha)$ represents the free-energy per particle of this short-ranged gas. 
Consequently, the prefactor in Eq. (\ref{eq:gap:typ:D}) for large $N$ behaves as
\begin{eqnarray} \label{prefactor}
N\frac{D_{\alpha}^2(\infty, N/2-1)}{D_{\alpha}(\infty, N)} \approx B A(\alpha)^2 \, N \;.
\end{eqnarray}         
Next, we consider the asymptotic large $N$ behavior of the function $F_\alpha$ that appears in the integrand in Eq. (\ref{eq:gap:typ:D}). Recalling that $F_\alpha(x,M)$ is the cumulative
distribution of the position of the rightmost particle in a gas of $M$ particles, one would expect that in the limit of large $M$, the function $F_{\alpha}(x, M)$ approaches its limiting form $F_{\alpha}(x)$ that is independent of $M$ \cite{Dhar2017}. In addition, this limiting form satisfies the non-linear eigenvalue equation (\ref{eq:intro:eigenvalue_eq}) where the same constant $A(\alpha)$ appears as the eigenvalue. Furthermore, $F_\alpha(x)$ has the asymptotic behaviors given in Eq. (\ref{asympt_Fa}). Setting further $x = \epsilon_{N/2+1}$ and $y = \epsilon_{N/2}$, we then get for large $N$
            \begin{align}\label{eq:gap:typ:E}
            \mathcal{P}_{\rm gap, bulk}(g, N)\approx N\, B [A(\alpha)]^{2}
            \int_{-\infty}^{\infty}dx\int_{-\infty}^{\infty}dye^{-\frac{1}{2}(x^2+y^2)}\\ 
            \delta(-x-y-g\,N+4\alpha)\theta(-x-y+4\alpha)
            F_{\alpha}(y+4\alpha)
            F_{\alpha}(x+4\alpha)\;.\nonumber
        \end{align}
This has clearly the scaling form 
\bea \label{eq:gap:typ:final2}
 \mathcal{P}_{\rm bulk, gap}(g, N) \approx N\,H_{\alpha}(gN) \;,
\eea        
with the scaling function
\bea \label{H1}
H_\alpha(z) =  B [A(\alpha)]^{2}
            \int_{-\infty}^{\infty}dx\int_{-\infty}^{\infty}dye^{-\frac{1}{2}(x^2+y^2)}\\ 
            \times \delta(-x-y-z+4\alpha)\theta(-x-y+4\alpha)
            F_{\alpha}(y+4\alpha)
            F_{\alpha}(x+4\alpha)\;.\nonumber
\eea
Note that for $z<0$ the ``delta-constraint'' and the ``theta-constraint'' in Eq. (\ref{H1}) can not satisfied simultaneously, implying that $H_\alpha(z) = 0$ for $z<0$. Hence the scaling function is supported on the positive semi-axis $[0, +\infty)$, which is expected since the positions are already ordered. For $z>0$, once the ``delta-constraint'' is satisfied, the ``theta-constraint'' is automatically satisfied. Hence we can get rid of the theta-function in Eq. (\ref{H1}). Performing the integral over $x$, we obtain  
 \be \label{H2}
 H_{\alpha}(z)= \theta(z) B\, A^2(\alpha) \int_{-\infty}^{\infty} dy F_{\alpha}(y+4\alpha) F_{\alpha}(8 \alpha -y-z) e^{-y^2/2-(y+z-4\alpha)^2/2}    \;,      
\ee
as announced in Eq. (\ref{eq:intro:gap_scaling_f}). Note that $A(\alpha)$ is already fixed from the eigenvalue equation (\ref{eq:intro:eigenvalue_eq}). Since the gap distribution in Eq. (\ref{eq:gap:typ:final2}) has to be normalised to one, one finds that, indeed, $B$ is the overall normalization constant and is of order $O(1)$, as assumed a priori. 

\begin{figure}
            \centering
            \includegraphics[width=0.7\linewidth]{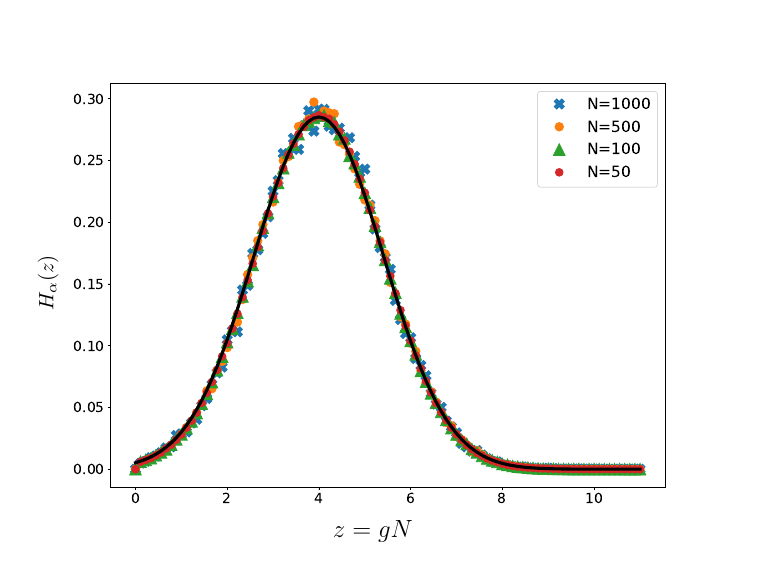}
            \caption{We measured the gap PDF ${\cal P}_{\rm gap, bulk}(g,N)$ in Monte-Carlo simulations for $N=50, 100, 500$ and $N=1000$, for fixed $\alpha=1$. The data collapse in a scaling form  
            ${\cal P}_{\rm gap, bulk}(g,N) \approx H_{\alpha=1}(g\,N)$ as predicted by the theory [see Eq. (\ref{eq:gap:typ:final2})]. The numerically obtained scaling function $H_{\alpha}(z)$ (shown with symbols) is in excellent agreement with the theoretical prediction in Eq. (\ref{H2}) with $\alpha = 1$ (solid line).}
            \label{fig:gap_typ:both}
        \end{figure}

We now derive the asymptotic behavior of the scaling function $H_\alpha(z)$ in Eq. (\ref{H2}). When $z \to 0$, it approaches a constant value. On the other hand, when $z \to +\infty$, it
decays rapidly. To derive this large $z$ behavior, we first rescale $y  = z\, u$ in the integral in Eq. (\ref{H2}). For large $z$, the integral reduces to (up to pre-exponential factors)
\bea \label{H_asympt1}
H_\alpha(z) \propto \int_{-\infty}^\infty du\, e^{-\frac{z^2}{2}(u^2+(u+1)^2)} F_\alpha(zu)\,F_\alpha(-(u+1)z) \;.
\eea
We now divide the integral over $u$ into three regions: (1) $u<-1$, (2) $-1 < u < 0$ and (3) $u>0$ and also recall the asymptotic behavior of $F_\alpha(x)$ as $x \to \pm \infty$ which reads
\bea \label{asymptF2}
F_\alpha(x) \approx e^{-\frac{|x|^3}{24\alpha}} \quad, \quad x \to - \infty \quad {\rm and} \quad F_\alpha(x) \approx 1 \quad, \quad x \to +\infty \;.
\eea
For the regime $u<-1$, the argument $-(u+1)z$ of the third factor in (\ref{H_asympt1}) approches to $+ \infty$ and hence the third term approaches $1$. In contrast, the argument of the second term $z\,u$ goes to $-\infty$ as $z \to \infty$. Hence the second term behaves as $\approx e^{- z^3 |u|^3/(24 \alpha)}$. The first term is subleading and behaves as $e^{-O(z^2)}$. Hence the contribution to the integral from the region $u \in (-\infty, -1]$, to leading order, is dominated by the upper limit of this region $u=-1$, and behaves as $I_1 \approx e^{-z^3/(24 \alpha)}$. Similarly, the contribution from the third region $u>0$ can be shown to be dominated by the lower edge at $u=0$ and it behaves for large $z$ as $I_3 \approx e^{-z^3/(24 \alpha)}$. The second region $-1<u<0$ is more interesting where the second and the third factors are of the same order. In this case the arguments $u\,z$ of the second factor and the argument $-(1+u)z$ of the third factor both tend to $- \infty$ as $z \to \infty$. Hence, using Eq. (\ref{asymptF2}), one finds that the integrand, to leading order for large $z$, behaves as $e^{-z^3((u+1)^3-u^3)/(24 \alpha)}$. We can now analyse the integral by a saddle-point method, where the minimum of $(u+1)^3-u^3)$ occurs at $u^*=-1/2$. Evaluating the integrand at the saddle point yields $I_2 \approx e^{- z^3/(96 \alpha)}$. Clearly $I_2 \gg I_1, I_3$ for large $z$. Hence, we obtain the asymptotic behavior of the bulk gap scaling function 
\bea \label{H_asympt2}
H_\alpha(z) \sim e^{- \frac{z^3}{96 \alpha} + o(z^3)} \;,
\eea
as announced in Eq. (\ref{asympt_H}) in the introduction. In Fig.~\ref{fig:gap_typ:both}, we tested the prediction of the scaling form in Eqs. (\ref{eq:gap:typ:final2}) and (\ref{H2}) against Monte-Carlo simulations, finding excellent agreement.

\section{Full counting statistics}\label{sec:FCS}

In this section, we are interested in the second important observable, namely the full counting statistics (FCS), i.e., the statistics of the number of particles $N_I$ in a given interval $I$ in the $1d$OCP. For a given configuration of charges with positions $\{x_i\}$, we can express this number of particles as
\begin{equation}
    N_I = \sum_{l=1}^{N}\mathbb{I}_I(x_l) \;,
    \label{eq:FCS:NI}
\end{equation}
where $\mathbb{I}_I(x)$ is a binary indicator function which takes the value $\mathbb{I}_I(x)=1$ if $x \in I$ and $\mathbb{I}_I(x)=0$ otherwise. Therefore, for any interval $I$, the random variable $N_I$ takes positive values $N_I = 0, 1, 2, \cdots, N$. The interval $I$ can be chosen either in the bulk or at the edges. For example, when $I$ is a semi-infinite interval, e.g., $I = [0, + \infty)$ it corresponds to the total number of charges on the positive half-axis and is often referred
to as the ``index''. The distribution of the index has been computed both for the log-gas \cite{Index1, Index2} as well as for the $1d$OCP \cite{Dhar2018}. Here, we consider the case where the interval is deep inside the bulk. While this bulk FCS has been well studied for the log-gas \cite{CL1995,FS1995,Marino2014,Marino2016}, to the best of our knowledge, it has not been studied for the $1d$OCP. Here we provide a detailed study of this bulk FCS. For simplicity, we choose here the interval to be symmetric around the origin, i.e. $I = [-L,+L]$, with $L \leq 2\alpha$ and $2\alpha - L = O(1)$. We just recall that $2 \alpha$ is the location of the right edge of the flat equilibrium density, which is supported on $[-2\alpha,2 \alpha]$ [see Eq. (\ref{eq:in:one})]. Note that   
our results detailed here for $I = [-L,+L]$ can be easily generalised to any interval in the bulk, not necessarily symmetric around the origin. 

The easiest statistics of $N_I$ is of course its mean value, which can be very easily computed. Taking average of Eq. (\ref{eq:FCS:NI}), one gets
\begin{equation}
    \bar{N}_I = N \,\int_{-L}^{+L} \tilde \rho(x) \, dx =\frac{L}{2\alpha}N \;,
    \label{eq:intro:average_FCS_2}
\end{equation}
where we used the equilibrium flat density $\tilde \rho(x)$ in Eq. (\ref{eq:in:one}). This random variable $N_I$ will of course fluctuate around this mean value from sample to sample. We will see below that the typical size of these fluctuations is of order $O(1)$, while the fluctuations of order $O(N)$ are atypical. As in the case of the bulk gap in the previous section, we will first compute the distribution of atypically large fluctuations of $N_I$, using a Coulomb gas method, which gives access to the large deviation regime. As we will see, this computation is relatively simpler than the computation of the distribution of the typical fluctuations, that is presented afterwards.

%However, we are interested in the full probability density defined as
%\begin{equation}
%    \mathcal{P} (N_I, N) = \frac{N!}{Z_N}\int_{-\infty}^{\infty}\prod_{k=1}^{N}dx_k e^{-S[\{x_i\}]}\prod_{i=2}^{N}\theta(x_{j}-x_{j+1}) \delta[\kappa_I N - \sum_{l=1}^{N}\mathbb{I}_I(x_l)],
%    \label{eq:FCS:LD_distribution}
%\end{equation}
%where $Z_N$ represents the partition function of the 1dOCP. Same as in the calculation of the gap, we only consider ordered configuration and multiply the result by number of possible permutations $N!$. We start by exploring the atypically large deviations in the limit of large $N$.

    \subsection{Large deviations of the FCS}
    
The starting point is the expression for the full distribution of the FCS, to leading order for large $N$, given by
 \begin{eqnarray}
\hspace*{-2.3cm}    \mathcal{P}_{\rm  FCS} (N_I, N) &=& \frac{\int_{-\infty}^{\infty}dx_1 \ldots \int_{-\infty}^{\infty}dx_N \, e^{-N^3\, S[\{x_i\}]}\prod_{i=2}^{N}\theta(x_{j}-x_{j+1}) \delta\left(N_I - \sum_{l=1}^{N}\mathbb{I}_I(x_l)\right)}{\int_{-\infty}^{\infty}dx_1 \ldots \int_{-\infty}^{\infty}dx_N e^{- N^3\, S[\{x_i\}]}\prod_{i=2}^N\theta(x_{j}-x_{j-1})} \nonumber \\
    \label{eq:FCS:LD_distribution}
    &=& \frac{\tilde{\cal N}_N}{{\cal D}_N} \;,
\end{eqnarray}   
where the denominator ${\cal D}_N$ is the same as in Eq. (\ref{eq:gap:large:distribution}) for the gap distribution, while the numerator $\tilde{\cal N}_N$ differs from the numerator ${\cal N}_N$ in Eq. (\ref{eq:gap:large:distribution}) since the delta function constraints are different in the two cases. Here, the action $S[\{x_i \}] = \beta E[\{ x_i\}]/N^3$, as before, is given explicitly in Eq. (\ref{eq:gap:large:action}) to leading order for large $N$. 

As in the case of the gap, our goal is to evaluate the multiple integrals in $\tilde {\cal N}_N$ and ${\cal D}_N$ by a saddle point method for large $N$. The denominator was already computed in Eq.~(\ref{DN}), for which we recall that the saddle point density was the flat density $\tilde \rho(x) = 1/(4 \alpha)$ for $-2 \alpha \leq x \leq 2 \alpha$. Hence, we just focus on the numerator here. Since, here, we are interested in the large fluctuations of $N_I$, of order ${O}(N)$, we first set $N_I = \kappa_I N$ where $0 \leq \kappa_I \leq 1$ is the fraction of charges in the interval $I$. We have seen from Eq. (\ref{eq:intro:average_FCS_2}) that the average value is $\bar \kappa = \bar{N_I}/N = L/(2 \alpha)$ in the large $N$ limit. We now want to evaluate the numerator for $\kappa_I > \bar{\kappa}$ and $\kappa_I < \bar{\kappa}$ separately.

\begin{figure}
                \centering
                \includegraphics[width=\linewidth]{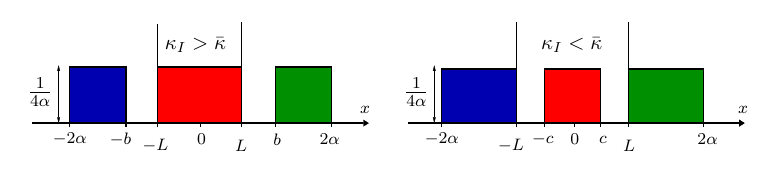}
                \caption{{\bf Left:} Schematic depiction of the ansatz for the saddle point density in Eq. (\ref{eq:FCS:ansatz_overp}) for the overpopulated regime, i.e. $\kappa_I > \bar{\kappa}$. {\bf Right:} Schematic depiction of the ansatz for the saddle point density in Eq. (\ref{eq:FCS:ansatz_underp}) for the underpopulated regime, i.e.~$\kappa_I < \bar{\kappa}$. The vertical lines at the $\pm L$ in both the figures represent a cartoon of the delta-peaks.}
                \label{fig:FCS:ansatz}
            \end{figure}

 \subsubsection{Overpopulated regime $\kappa_I>\bar{\kappa}_I$.} 
\begin{figure}
        \centering
        \includegraphics[width=\linewidth]{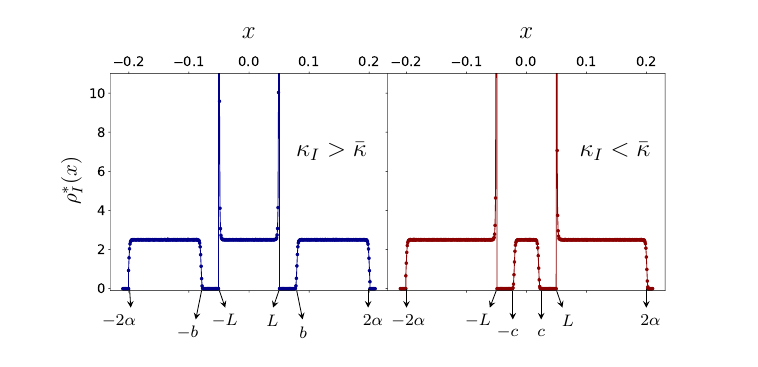}
        \caption{Monte-Carlo simulations for the charge density with $N=1000$ particles, $\alpha =0.1$ and $L=0.05$. In this case ${\bar{\kappa}}=L/(2\alpha)=0.25$. {\bf Left:} Overpopulated regime $\kappa_I = 0.4 > \bar{\kappa}=0.25$, which confirms the ansatz in Eq. (\ref{eq:FCS:ansatz_overp}). {\bf Right:} Underpopulated regime $\kappa_I = 0.1 < \bar{\kappa}=0.25$, confirming the ansatz in Eq. (\ref{eq:FCS:ansatz_underp}).}
        \label{fig:FCS:MC_simulation}
    \end{figure}

In this regime, the configurations that contribute to the multiple integral $\tilde{\cal N}_N$ in (\ref{eq:FCS:LD_distribution}) contain a fraction
 $\kappa_I > \bar{\kappa}$, i.e., more than the average fraction of charges at equilibrium are pushed inside the interval $I = [-L,+L]$. Since the total number of charges $N$ is conserved, this leads to a depletion of charges symmetrically for both intervals $[-2 \alpha, -L]$ and $[L,2 \alpha]$. Thus we expect that the saddle point density will consist of three disjoint intervals, with the middle one overpopulated compared to the equilibrium average. From our general discussion about the behavior of the 
 Coulomb gas in a harmonic trap [see below Eq. (\ref{Ystar})], we would expect that the transfer of extra-charges from the two side-intervals into the middle one will leave the bulk density everywhere unaffected, i.e., flat with height $1/(4\alpha)$ over the three intervals $[-2 \alpha, -b]$, $[-L,+L]$ and $[+b,2 \alpha]$. The extra charges that move into the middle interval accumulate symmetrically as delta-peaks at the two edges $-L$ and $+L$ (see the left panel of Fig. \ref{fig:FCS:MC_simulation}). Note that we have assumed that the two outer edges $\pm 2 \alpha$ of the $1d$OCP remain unperturbed in the presence of the constraint that the middle interval has $\kappa_I > \bar \kappa$ fraction of charges. This leads us to the ansatz for the saddle point in the presence of the delta constraint in the numerator $\tilde{\cal N}_N$ in Eq. (\ref{eq:FCS:LD_distribution})
\begin{align}
                \rho_I^*(x)^=\begin{cases}
                \frac{1}{4\alpha}, \quad &-2\alpha \leq x \leq -b,\\
                \frac{1}{4\alpha}+B\delta(x+L)+B\delta(x-L), \quad &-L\leq x \leq L \;,\\
                \frac{1}{4\alpha}, \quad &b\leq x\leq 2\alpha \;,
                 \end{cases}
            \label{eq:FCS:ansatz_overp}
            \end{align}
characterised by the two unknown parameters $b\leq L$ and $B>0$. For a schematic depiction of this saddle point density in the overpopulated regime, see the left panel of Fig. \ref{fig:FCS:ansatz}. 
To fix these parameters, we can use two sum rules: 
 \begin{itemize}
                \item The fraction of particles in the middle block (shown in red in Fig. \ref{fig:FCS:ansatz}) is $\kappa_I$, i.e., $\int_{-L}^{+L}  \rho^*_I(x)\, dx = \kappa_I$. Substituting the ansatz (\ref{eq:FCS:ansatz_overp}) in this integral gives
                     \begin{equation}
                          2\left(B +\frac{L}{4\alpha} \right)=\kappa_I \;.
                          \label{eq:FCS:overp_c1}
                      \end{equation}
               \item The fraction of particles that get depleted from the interval $[+L,+b]$ (or equivalently from $[-b,-L]$) piles up as a delta peak at the right edge (equivalently at the left edge) of the middle interval. This leads to
                  \begin{equation}
                     \frac{b-L}{4\alpha}=B \;.
                     \label{eq:FCS:overp_c2}
                  \end{equation}
                          \end{itemize}
Solving for the two unknowns $B$ and $b$ we get 
 \begin{eqnarray}
            B = \frac{2\alpha\kappa_I -L}{4\alpha} \quad, \quad b = 2\alpha\kappa_I \;, \label{eq:FCS:overp_solution}
 \end{eqnarray}
which characterizes the saddle point density $\rho^*_I(x)$ in Eq. (\ref{eq:FCS:ansatz_overp}) completely.  

Note that, unlike in the case of the gap distribution in Section \ref{sec:bulkgap}, here we did not minimize the effective action associated to the numerator $\tilde{\cal N}_N$ in 
Eq. (\ref{eq:FCS:LD_distribution}) by assuming arbitrary parameters for the height $h$ in the bulk 
and the outer edges $\pm a$. But, as discussed previously below Eq. (\ref{Ystar}), if we had kept these parameters arbitrary and minimise this effective action, we would indeed get that $h = 1/(4 \alpha)$ and $a = 2 \alpha$. One can check this explicitly but we do not repeat the calculation here, since it is exactly similar to that of the gap distribution in Section \ref{sec:bulkgap} and is a general feature of the $1d$OCP. Hence, very much to our convenience, we can write explicitly the saddle point density 
just by symmetry arguments and sum rules, without having to minimize the effective action. 
associated to numerator $\tilde{\cal N}_N$ in Eq. (\ref{eq:FCS:LD_distribution}). As argued above, this density is guaranteed to be the minimum of the effective action, quite generally for a Coulomb gas in a harmonic potential. Our Monte-Carlo simulations confirm this as shown in Fig.~\ref{fig:FCS:MC_simulation} (left panel).

We now evaluate the saddle point action by substituting $\rho_I^*(x)$ from Eq. (\ref{eq:FCS:ansatz_overp}) into the action in Eq. (\ref{eq:gap:large:action}). One can check easily that the contribution to the action from the bulk part of $\rho_I^*(x)$, for large $N$, vanishes exactly and the only nonvanishing contributions come from the two delta peaks. To evaluate the contribution from the left delta peak, we substitute $x_ i = -L$ for $i = (1 -\kappa_I)N/2$ up to $i = (1- L/(2 \alpha))N/2$ in Eq. (\ref{eq:gap:large:action}). Similarly, we do the same for the right delta peak which is exactly the same as the left one, by symmetry. Replacing the sum by integral in the large $N$ limit, we get  
\bea \label{Sdelta}
S^* &=& \int_{(1-\kappa_I)/2}^{(1-L/(2 \alpha))/2} \left(-L - 4 \alpha \,z + 2 \alpha \right)^2 \, dz - \frac{2 \alpha^2}{3}  \nonumber \\
&=& \frac{(2 \alpha \kappa_I-L)^3}{12 \alpha} - \frac{2 \alpha^2}{3} \;.
\eea  
Therefore the numerator in Eq. (\ref{eq:FCS:LD_distribution}) is given, for large $N$, by $\tilde {\cal N}_N \sim e^{-N^3\,S^*}$. Using the result for the denominator in Eq. (\ref{DN}) and taking the ratio in Eq. (\ref{eq:FCS:LD_distribution}), we get
\bea \label{P_FCS_2}
{\cal P}_{\rm FCS}(N_I = \kappa_I\,N,N) \sim e^{-N^3 \psi(\kappa_I)} \quad \;, \quad \kappa_I > \bar{\kappa} = \frac{L}{2\alpha} \;,
\eea  
where the rate function $\psi(z)$ is given by
\bea \label{psi_text2}
\quad \psi(z) = \frac{2}{3}\alpha^2\left(z - \frac{L}{2\alpha}\right)^3 \quad, \quad {\rm for} \quad \quad z > L/(2 \alpha) \;.
\eea

 \subsubsection{Underpopulated regime, $\kappa_I< \bar{\kappa}$}

     In this case, since $\kappa_I< \bar{\kappa}$, the configurations that contribute to the integral in ${\tilde {\cal N}}_N$ in (\ref{eq:FCS:LD_distribution}), will have less charges in the interval $[-L,+L]$ compared to the equilibrium configuration. These depleted charges will get distributed symmetrically on either side of $[-L,+L]$. From the general discussion on the $1d$OCP from the previous section, these extra charges will pile up as delta peaks of equal magnitude at $\pm L$, while keeping the height at $1/(4 \alpha)$ and the two edges at $\pm 2 \alpha$, as in the equilibrium configuration. Consequently, the support of the middle interval, shown by the red block in Fig. \ref{fig:FCS:ansatz} will also shrink to $[-c,+c]$ with $c \leq L$. This leads to the ansatz for the saddle point density   
 \begin{align}
                \rho_I^*(x)^=\begin{cases}
                \frac{1}{4\alpha} + C\delta(x+L) \quad &-2\alpha \leq x \leq -L,\\
                \frac{1}{4\alpha}, \quad &-c\leq x \leq +c \;,\\
                \frac{1}{4\alpha}+C\delta(x-L), \quad &L\leq x\leq 2\alpha \;,
                 \end{cases}
            \label{eq:FCS:ansatz_underp}
            \end{align}
characterised by the two unknown parameters $c\leq L$ and $C>0$. For a schematic depiction of this saddle point density in the underpopulated regime, see the right panel of Fig. \ref{fig:FCS:ansatz}. To fix these two parameters, we use the two sum rules
\bea\label{sum_ruleunderp}
\int_{-c}^{+c} \rho_I^*(x)\,dx = \kappa_I \quad {\rm and} \quad \int_{L}^{2 \alpha} \rho_I^*(x)\,dx = \frac{1-\kappa_I}{2} \;.
\eea
Substituting the ansatz (\ref{eq:FCS:ansatz_underp}) in these integrals (\ref{sum_ruleunderp}) fixes the two parameters
\bea\label{sum_ruleunderp2}
c = 2 \alpha \kappa_I \quad {\rm and} \quad C = \frac{L-2\alpha \kappa_I}{4\alpha} \;.
\eea
As in the case $\kappa_I > \bar \kappa$, one can verify that this ansatz (\ref{eq:FCS:ansatz_underp}) indeed is the saddle point density, i.e., it minimises the effective
action associated to the multiple integral for $\tilde{\cal N}_N$ in Eq. (\ref{eq:FCS:LD_distribution}). Once again, our Monte-Carlo simulations confirm this ansatz 
as shown in Fig.~\ref{fig:FCS:MC_simulation} (right panel). 

To evaluate the saddle point action, we substitute this ansatz (\ref{eq:FCS:ansatz_underp}) in Eq. (\ref{eq:gap:large:action}). The contribution from the bulk again vanishes, as it should, since
the height in the bulk remains unchanged at $1/(4\alpha)$, as in the equilibrium configuration. Thus, the only nonvanishing contribution to the saddle point action comes from the two delta peaks. This gives
\bea \label{Star_underpop}
S^* = \frac{1}{N} \sum_{i = \frac{N}{4\alpha}(2\alpha-L)}^{\frac{N}{4\alpha}(2\alpha-L)+CN} \left(-L - \frac{2\alpha}{N}(2i-N-1)\right)^2 - \frac{2}{3}\alpha^2 \;.
\eea   
In arriving at this result, we substituted $x_i = -L$ for all the charges in the left delta peak (the right delta peak gives exactly the same contribution, leading to the absence of the overall factor $1/2$ i the first term of (\ref{Star_underpop}). The lower and the upper indices in the summation in the first term can be easily obtained. For instance, the lower index is obtained by setting $-L = 4 \alpha i/N-2 \alpha$, corresponding to the index of the last particle in the left block. Replacing the sum by an integral in the large $N$ limit, and perform the integral explicitly gives
\bea \label{Star_underpop2}
S^* = \frac{1}{12 \alpha} (4 \alpha C)^3- \frac{2}{3}\alpha^2  = \frac{1}{12 \alpha}\left( L - 2 \alpha\, \kappa_I \right)^3- \frac{2}{3}\alpha^2\;.
\eea
where we used the value of $C$ from Eq. (\ref{sum_ruleunderp2}). Therefore the numerator in Eq. (\ref{eq:FCS:LD_distribution}) is given, for large $N$, by $\tilde{\cal N}_N \sim e^{-N^3\,S^*}$. Using the result for the denominator in Eq. (\ref{DN}) and taking the ratio in Eq. (\ref{eq:FCS:LD_distribution}), we get
\bea \label{P_FCS_3}
{\cal P}_{\rm FCS}(N_I = \kappa_I\,N,N) \sim e^{-N^3 \psi(\kappa_I)} \quad \;, \quad \kappa_I < \bar{\kappa} = \frac{L}{2\alpha} \;,
\eea  
where the rate function $\psi(z)$ is given by
\bea \label{psi_text3}
\quad \psi(z) = \frac{2}{3}\alpha^2\left(\frac{L}{2\alpha}-z\right)^3 \quad, \quad {\rm for} \quad \quad z < L/(2 \alpha) \;.
\eea

Having obtained the distribution of the atypical large fluctuations for both the overpopulated (\ref{psi_text2}) and the underpopulated (\ref{psi_text3}) regimes, we see that we can
combine the two results into a single large deviation form 
\bea \label{P_FCS_22}
{\cal P}_{\rm FCS}(N_I = \kappa_I\,N,N) \sim e^{-N^3 \psi(\kappa_I)} \quad \; {\rm where} \quad \psi(z) =   \frac{2}{3}\alpha^2\left|z - \frac{L}{2\alpha}\right|^3 \;.
\eea  
Let us remark that this rate function $\psi(z)$ is singular at its minimum $z = L/(2\alpha)$ with a discontinuous third derivative. Usually, in systems with short-ranged correlations, 
one would expect, from the central limit theorem, that the distribution of the FCS is Gaussian near its peak, implying that the associated rate functions describing large deviations should behave quadratically near its minimum. Remarkably, it turns out that, here, the rate function is {\it singular} at its minimum, thus violating the central limit theorem. This clearly happens due to the long-range nature of the repulsive interactions.

      \subsection{Typical fluctuations of the FCS}
     In this section we compute the typical fluctuations of the number of particles in $[-L,+L]$ around the mean. As we show below, the distribution of the typical fluctuations can be characterised by the scaling form given in Eq. (\ref{eq:int:FCS1}), namely
\bea \label{scaling_form}
{\cal P}_{FCS}(N_I, N) \sim 2 \alpha U_\alpha(2 \alpha(N_I - \bar{N}_I))  \;,
\eea     
where the scaling function $U_\alpha(z)$ can be expressed also in terms of the function $F_\alpha(x)$ that appears in the computation for the distribution of the gap in Section \ref{Sec:typgap}. 
     
\begin{figure}
        \centering
        \includegraphics[width=12cm]{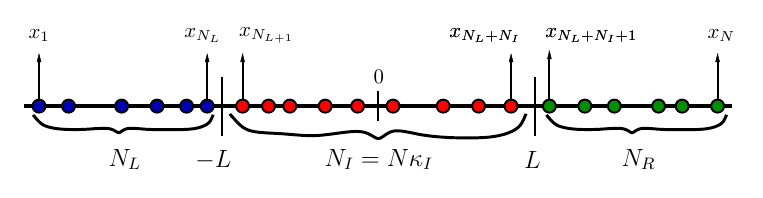}
        \caption{Schematic representation of a configuration of charges where the charges inside the interval $[-L,+L]$ are shown by red filled circles. Those to the left of $-L$ are shown in blue, while those to the right of $+L$ are shown by green. We denote the number of charges in the three intervals respectively by $N_L, N_I$ and $N_R$. The location of the particles that are at the edges of these intervals are also marked.}
        \label{fig:FCS:typ:picture}
    \end{figure}

We are interested in the probability distribution that the number of particles in the interval $I=[-L, L]$ is equal to $N\kappa_I$ when the total number of particles is $N$. To compute this distribution, it is convenient to first consider the joint distribution ${\cal P}_{\rm joint}(N_L, N_R|N)$ of $N_L$ and $N_R$, denoting respectively the number of charges to the left of $-L$ and to the right of $+L$. The number of particles $N_I$ in the interval $I$ is obtained from the sum rule, $N_I = N-N_L-N_R$. We can then obtain the marginal distribution ${\cal P}_{FCS}(N_I,N)$ by summing over $N_L$ 
\bea \label{marginal}
{\cal P}_{FCS}(N_I, N) = \sum_{N_L=0}^N {\cal P}_{\rm joint}(N_L, N_R=N-N_L-N_I|N) \;.
\eea
For an ordered configuration with positions $x_1<x_2 < \cdots < x_N$, this joint distribution can be expressed as a multiple integral
     \begin{align}
        &\mathcal{P}_{\rm joint}(N_L, N_R| N) =  \frac{N!}{Z_N}\int_{-\infty}^{\infty}\prod_{k=1}^{N}dx_k e^{-\beta\,E[\{x_i\}]}\prod_{k=2}^{N}\theta(x_k-x_{k-1}) \nonumber\\ 
        & \times \theta(-L-x_{N_L})\theta(L+x_{N_L+1})\theta(L-x_{N_L+N_I})\theta(x_{N_L+N_I+1}-L)\, \;,\nonumber \\
    \label{eq:FCS:typ:probability}
\end{align}
where $\beta E[\{x_i\}]$ is given in Eq. (\ref{eq:in:energy_simpler}), $Z_N$ in Eq. (\ref{eq:int:Zn}) and $N_I = N-N_L-N_R$. The product of theta functions ensures that the particles are ordered. The theta function $\theta(-L-x_{N_L})$ enforces the number of particles to the left of $-L$ to be exactly $N_L$ (which is equivalent to say that $x_{N_L} \leq -L$ since the postions are ordered). The product of the two theta functions $\theta(L+x_{N_L+1})\theta(L-x_{N_L+N_I})$ ensures that the number of particles in the interval $[-L,+L]$ is $N_I$. The last theta  function $\theta(x_{N_L+N_I+1}-L)$ guarantees that the number of particles to the right of $+L$ is exactly $N_R = N-N_L-N_I$ (see Fig. \ref{fig:FCS:typ:picture} for a schematic representation). 

    It is convenient to split this multiple integral in (\ref{eq:FCS:typ:probability}) into three blocks from left to right
    \begin{eqnarray}
 &&  \hspace*{-2.cm}    \mathcal{P}_{\rm joint}(N_L, N_R| N) = \frac{N!}{Z_N}\mathcal{I}_1\mathcal{I}_2\mathcal{I}_3\nonumber \\
        && \hspace*{-2.cm}=\frac{N!}{Z_N} \left[ \int_{-\infty}^{\infty}\prod_{k=1}^{N_L}dx_k e^{-\beta E[\{x_i\}]}\prod_{k=2}^{N_L}\theta(x_k-x_{k-1})\theta(-L-x_{N_L})\right] \\
        &&\hspace*{-2.cm}\times \left[ \int_{-\infty}^{\infty}\prod_{k=N_L+1}^{N_L+N_I}dx_k e^{-\beta E[\{x_i\}]}\prod_{k=N_L+2}^{N_L+N_I}\theta(x_k-x_{k-1})\theta(x_{N_L+1}+L)\theta(L-x_{N_L+N_I}) \right]\nonumber \\
       &&\hspace*{-2.cm}\times \left[ \int_{-\infty}^{\infty}\prod_{k=N_L+N_I+1}^{N}dx_k e^{-\beta E[\{x_i\}]}\prod_{k=N_L+N_I+2}^{N}\theta(x_k-x_{k-1})\theta(x_{N_L+N_I+1}-L)\right].\nonumber
    \label{eq:FCS:typ:three_int}
    \end{eqnarray}
While this is an exact expression valid for any finite $N$, it is hard to extract any explicit form for the joint distribution, due to the presence of too many constraints enforced by the theta functions. However, as we show below, for large $N$ one can use an alternative method, without having to evaluate these constrained multiple integrals explicitly. 

 \begin{figure}
        \centering
        \includegraphics[width=0.7\linewidth]{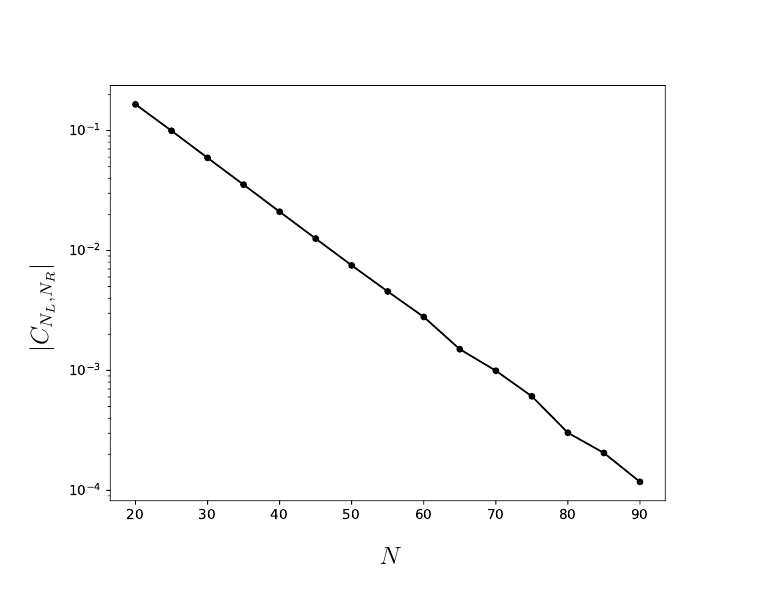}
        \caption{The absolute value of the covariance $|C_{N_L,N_R}(N)|$ in Eq. (\ref{eq:FCS:typ:correlation}) between $N_L$ and $N_R$, is plotted on a semi-log plot as a function of $N$, for the choice of parameters $\alpha = 0.1$ and $L=0.05$. The straight line with a negative slope confirms the exponential decay of the covariance function with increasing $N$.}
        \label{fig:FCS:typ:correlations}
    \end{figure}
The main difficulty in computing the joint distribution of $N_L$ and $N_R$ is due to the fact that they are correlated for any finite $N$. However, for large $N$, a simplification arises due to the fact that $N_L$ and $N_R$ get essentially uncorrelated. This can be seen from the expression of the energy in Eq. (\ref{eq:in:energy_simpler}), which
shows that the fluctuations of the positions around the ``crystal configuration'' (where the charges are equivalent-spaced with a distance $4\alpha/N$) are essentially localised in space. The long-range nature of the interactions is responsible for creating the ``crystal configuration'' but the fluctuations around this configuration essentially are short ranged. Hence, if we want to increase $N_L$ by order $O(1)$, by moving some charges from $[-L,+L]$ to the left of $-L$, it involves the energy cost, which is localised around $-L$ and hence will not affect $N_R$ at all, as long as $L \gg 4\alpha/N$. Even when the fluctuations in $N_L$ is large, i.e., of order $O(N)$, we have seen in the computation of the large deviation function that the dominant contribution to the change in energy comes from the delta-peak at $-L$, which again shows that the fluctuations of $N_L$ are
uncorrelated with $N_R$. Due to the short-range nature of the fluctuations around the crystal configuration, we expect that the covariance function,   
\begin{equation}
    C_{N_L,N_R}(N) =  \overline{N_L N_R} - \bar{N}_L\bar{N}_R \;,
    \label{eq:FCS:typ:correlation} 
     \end{equation}
where $\bar N_L$ and $\bar N_R$ are the respective averages, decreases exponentially with increasing $N$. Numerical simulations indeed confirm this fact, see Fig. \ref{fig:FCS:typ:correlations}.
 
Since this covariance decays to zero for large $N$, one can approximate the joint distribution by the factorization
\bea \label{approx}
\mathcal{P}_{\rm joint}(N_L, N_R| N) \approx {\cal P}(N_L|N)\, {\cal P}(N_R|N) \;,
\eea
where ${\cal P}(N_L|N)$ and ${\cal P}(N_R|N)$ are the distributions of the number of particles to the left of $-L$ and to the right of $+L$ respectively. Substituting this approximation, valid for large $N$, in Eq. (\ref{marginal}), we can express the marginal distribution of $N_I$ as 
\bea \label{marginal_approx}
{\cal P}_{FCS}(N_I, N) \approx \sum_{N_L=0}^N {\cal P}(N_L|N)\, {\cal P}(N_R = N-N_L-N_I|N)  \;.
\eea
By symmetry, one would expect further that the sum will be dominated by configurations where $N_L = N_R = (N-N_I)/2$. Hence, we can further approximate the marginal distribution of $N_I$ by
\bea \label{marginal_approx2}
{\cal P}_{FCS}(N_I, N) &\approx& \tilde C \; {\cal P}\left(N_L = \frac{N-N_I}{2}\Big |N\right)\, {\cal P}\left(N_R = \frac{N-N_I}{2}\Big |N\right) \nonumber \\
&=&  \tilde C \, {\cal P}^2\left(N_L=\frac{N-N_I}{2}\Big |N\right) \;,
\eea
where $\tilde C$ is a constant entropy factor, independent of $N_I$. Therefore, to calculate the leading order behavior of the typical distribution of the FCS ${\cal P}_{FCS}(N_I, N)$ for large $N$, we just need to compute the distribution of the number of 
particles to the left of $-L$ in the $1d$OCP. We note that the typical distribution of the number of particles to the left of the origin ${\cal P}(N_-,N)$, known as the index distribution, was computed analytically for large $N$ recently in Ref. \cite{Dhar2018}, leading to the result
\bea \label{res_index}
{\cal P}(N_-,N) \approx 4\alpha f_\alpha(4 \alpha(N_- - \bar{N}_-)) \quad {\rm with} \quad  \bar{N}_- = \frac{N}{2} \;,
\eea
where the scaling function $f_\alpha(z)$ was computed explicitly to be
\bea \label{scal_f}
f_\alpha(z) = \frac{F_\alpha(z+2\alpha)\, F_\alpha(-z+2\alpha)}{\int_{-\infty}^\infty F_\alpha(z+2\alpha)\, F_\alpha(-z+2\alpha)\, dz} \;,
\eea
in terms of the same function $F_\alpha(x)$ representing the solution of Eq. (\ref{eq:intro:eigenvalue_eq}). The typical fluctuations of $N_-$ around its
mean value $\bar{N}_- = N/2$ are of order $O(1)$, as evident from the scaling form (\ref{res_index}). In our problem, for the distribution of the FCS in Eq. (\ref{marginal_approx2}),
we need instead to compute the distribution of the number of particles to the left of $-L$ (and not the origin as in the index problem). However, we can follow exactly the same steps
as in Ref. \cite{Dhar2018}  (not repeated here), see also \cite{rojas18}, and we obtain 
\bea \label{PNL_index}
{\cal P}(N_L,N) \approx 4 \alpha f_\alpha(4\alpha(N_L - \bar{N}_L)) \quad {\rm where} \quad \bar{N}_L=\frac{2\alpha-L}{4\alpha}N \;.
\eea
Thus the only thing that changes from the computation of the index distribution is the fact that the center of the scaling function is now at $\bar N_L=\frac{2\alpha-L}{4\alpha}$, denoting
the average number of particles to the left of $-L$. Of course, for $L=0$, we recover the result for the index distribution. Using this result (\ref{PNL_index}) in Eq. (\ref{marginal_approx2}), we then obtain the FCS distribution for the typical fluctuations
\bea \label{FCS_inter1}
{\cal P}_{FCS}(N_I, N) \approx 16\alpha^2 \tilde C \left[ f_\alpha\left( -2 \alpha(N_I - \bar{N}_I)\right)\right]^2 \quad {\rm where} \quad \bar{N}_I = \frac{L}{2\alpha}\,N \:.
\eea
We can now fix the constant $\tilde C$ from the normalisation condition $\sum_{N_I} {\cal P}_{FCS}(N_I, N) = 1$. Consequently, we get the scaling form in (\ref{scaling_form}) for ${\cal P}_{FCS}(N_I, N)$, where the scaling function $U_\alpha(z)$, normalised to $1$, is given by
\bea \label{U_function}
U_\alpha(z) = \frac{F^2_\alpha(2\alpha+z) F_\alpha^2(2\alpha-z)}{\int_{-\infty}^\infty F^2_\alpha(2\alpha+z) F_\alpha^2(2\alpha-z)\, dz} \;.
\eea
 \begin{figure}[t]
        \centering
        \includegraphics[width=0.8\linewidth]{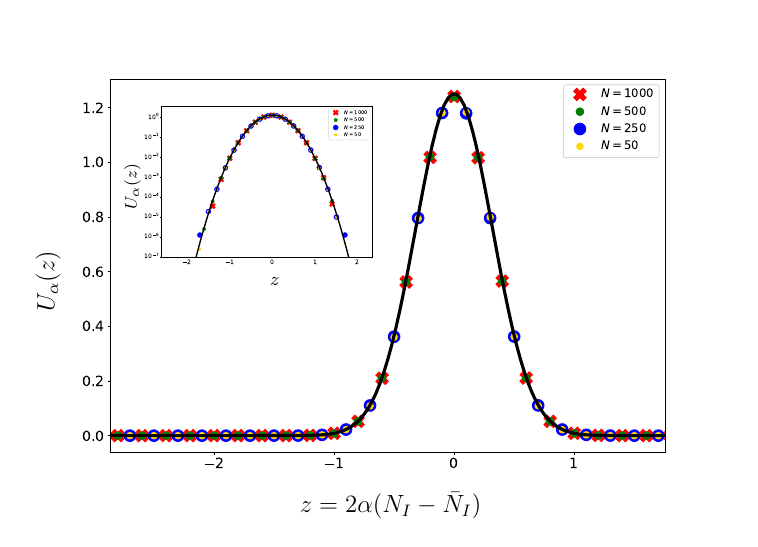}
        \caption{We verify in Monte-Carlo simulations that ${\cal P}_{\rm FCS}(N_I,N)$ indeed satisfies the scaling form in Eq. (\ref{scaling_form}). The data for four values of $N=50, 250, 500$ and $1000$ collapse onto a single scaling curve, as shown by the symbols. This is compared with the theoretical prediction for the scaling function $U_\alpha(z)$ in Eq. (\ref{U_function}) shown by the solid line. In Eq. (\ref{U_function}), the scaling function $U_\alpha(z)$ was evaluated by computing $F_\alpha(z)$ numerically from Eq. (\ref{eq:intro:eigenvalue_eq}). The value of the parameters are       $\alpha=0.1$ and $L=0.05$. The inset shows the same data in semi-log plot. The agreement is excellent.}
        \label{fig:FCS:typ:typical}
    \end{figure}
This scaling function is symmetric, $U_\alpha(z)=U_\alpha(-z)$. Furthermore, using the asymptotic behaviors of $F_\alpha(x)$ in Eq. (\ref{asympt_Fa}), we find that $U_\alpha(z)$ has a non-Gaussian tail for large $|z|$, namely
\bea \label{U_large_z}
U_\alpha(z) \approx e^{-\frac{|z|^3}{12\alpha}} \;.
\eea 
In Fig. \ref{fig:FCS:typ:typical}, we compare our analytical prediction for the scaling form (\ref{scaling_form}) with the scaling function $U_\alpha(z)$ given in (\ref{U_large_z}) 
with Monte-Carlo simulations, finding a good agreement.

\vspace*{0.5cm}   

\noindent We end this section with two remarks:

\vspace*{0.5cm}     
                     
\noindent {\bf Remark 1: Strongly hyper-uniform nature of the FCS fluctuations.} From our main result in Eq. (\ref{scaling_form}), we see that typical fluctuations of the FCS in an interval $[-L,+L]$
is of order $O(1)$, independent of both $N$ and $L$. It is interesting to compare this result to other particle systems in a harmonic well, both non-interacting as well as logarithmically 
repulsive (the so-called ``log-gas'' that appears in random matrix theory). In the completely noninteracting case, where the FCS has Poisson statistics, it is clear that the variance of 
$N_I$ is proportional to the average $\bar {N_I}$. In the log-gas case, it is also well known \cite{CL1995,FS1995,Marino2014,Marino2016} that the variance of the typical fluctuations of $N_I$ scales as $\log(\bar{N}_I)$. Thus the fluctuations get suppressed due to the long-range logarithmic repulsion. Such a gas with suppressed fluctuations is usually referred to as a {\it hyper-uniform} gas \cite{torquato}. In the $1d$OCP case, we see that the fluctuations are of order $O(1)$ and do not increase with $\bar{N}_I$. Thus the $1d$OCP gas is even more hyper-uniform compared to the log-gas.      

\vspace*{0.5cm}     
                     
\noindent {\bf Remark 2: The relation between the gap distribution and the FCS.} In this paper, we have studied two observables for the $1d$OCP: (i) the gap distribution in the bulk, i.e., the distribution of the distance $g_k = x_{k+1}-x_k$ between the $k$-th and $(k+1)$-th particles in the bulk. We have argued that this distribution does not depend on the particle label $k$, as long as $k$ belongs to the bulk and, hence, for convenience we have set $k=N/2$, i.e., the mid-gap. (ii) The probability distribution of the FCS, i.e., the distribution ${\cal P}_{FCS}(N_I, N)$ of the number of particles $N_I$ in the interval $[-L,+L]$. One naturally wonders whether there is any relation between these two observables.  The answer is a priori 'no', since in case (i), the gap is defined with respect to a fixed label attached to a particle, i.e., in the ``particle frame'', while in (ii) the interval $[-L,+L]$ is fixed in the ``lab frame''. To see this more precisely, we first consider the following exact relation
\bea \label{gap_FCS1}
{\cal P}_{FCS}(N_I = 0, N) = \sum_{N_L=0}^N {\rm Prob.}\left[x_{N_L}\leq -L, g_{N_L} \geq 2L \right] \;.
\eea  
This relation can be understood as follows. The event that the interval $[-L,+L]$ is empty is equivalent to the event that there are $N_L$ particles to the left of $-L$ and $N-N_L$ particles 
to the right of $+L$, i.e., the position $x_{N_L}$ of the $N_L$-th particle is less than $-L$ and the gap in front of it is bigger than $2L$. Finally, of course, we need to sum over all possible
values of $N_L=0,1, \cdots, N$. In general, the right hand side of Eq. (\ref{gap_FCS1}) involves the joint distribution of $x_{N_L}$ and the gap $g_{N_L}$ in front of it, which cannot be reduced to the marginal distribution of the gap only. Therefore, in general, there is no simple relation between the emptiness probability of the interval $[-L,+L]$ and the gap distribution. 
However, for large $N$, and for atypically large fluctuations of the gap, one can find an approximate relation connecting the two observables. To see this, we note that, in the large $N$ limit, and
when the gap is large, the sum on the right hand side in Eq. (\ref{gap_FCS1}) will be dominated by the ``saddle point type'' configuration as shown in the right panel of Fig. \ref{fig:FCS:MC_simulation} with $\kappa_I = 0$. In that case, the right hand side of Eq. (\ref{gap_FCS1}) coincides with the probability that the mid-gap $g_{N/2} = 2L$, i.e.,
\bea \label{gap_FCS2}
{\cal P}_{FCS}(N_I = 0, N)  \approx {\cal P}_{\rm gap, bulk}(g=2L,N) \;.
\eea
This relation is confirmed by our exact calculation of the large deviation probability associated with the FCS in Eq. (\ref{P_FCS_2}) and with the mid-gap in Eq. (\ref{eq:gap:large:ldf}). Indeed, by setting $g=2L$ in Eq. (\ref{eq:gap:large:ldf}), and $\kappa_I = 0$ in Eq. (\ref{eq:gap:large:ldf}), one can easily verify that both probabilities behave as $\approx e^{-N^3 L^3/(12\alpha)}$, thus confirming the relation in Eq. (\ref{gap_FCS2}).

\section{Conclusion}

In this paper, we have studied analytically two observables associated with the $1d$OCP: (i) the distribution of the gap between two consecutive particles in the bulk and 
(ii) the distribution of the number of particles $N_I$ in a fixed interval $I=[-L,+L]$ inside the bulk, the so-called FCS. For both observables, we have studied the distribution of the typical as well as atypical large fluctuations and shown that they are described by different functional forms. For the case of the gap, we have shown that the distribution of the typical fluctuations are described by the scaling form ${\cal P}_{\rm gap, bulk}(g,N) \sim N H_\alpha(g\,N)$, where $\alpha$ is the interaction coupling and the scaling function $H_\alpha(z)$ is given explicitly in Eq. (\ref{eq:intro:gap_scaling_f}). This scaling behavior indicates that the typical size of the fluctuations of the gap is of order $O(1/N)$. For fluctuations of order $O(1)$, the distribution is described by the large deviation form ${\cal P}_{\rm gap, bulk}(g,N) \sim e^{-N^3 \psi_{\rm bulk}(g)}$, where the rate function $\psi_{\rm bulk}(g) = g^3/(96\alpha)$ with $g\geq 0$. Similarly, for the FCS, we show that the distribution of the typical fluctuations of $N_I$ is described by the scaling form ${\cal P}_{\rm FCS}(N_I,N) \sim 2\alpha \, U_\alpha[2 \alpha(N_I - \bar{N}_I)]$, where $\bar{N}_I = L\,N/(2 \alpha)$ is the average value of $N_I$ and the scaling function $U_\alpha(z)$ is given in Eq. (\ref{eq:intro:FCS_typical}). Thus the typical fluctuations of $N_I$ around its mean are of order $O(1)$. On the other hand, the atypical large fluctuations of $N_I$, of order $O(N)$, are described the large deviation form ${\cal P}_{\rm FCS}(N_I,N) \sim e^{-N^3 \Phi(N_I/N)}$ where $\Phi(z) = (2\alpha^2/3) |z-L/(2\alpha)|^3$. We note that both the scaling functions describing typical fluctuations of the gap and the FCS, namely $H_\alpha(z)$ and $U_\alpha(z)$ are expressed in terms
of the same function $F_\alpha(x)$ that appears as the solution of the nonlinear eigenvalue problem in Eq. (\ref{eq:intro:eigenvalue_eq}).   

We note that the distribution of the typical gap in the bulk was recently studied numerically in Ref. \cite{Santra2021}, for the general Riesz gas with various indices $k>-2$ [see Eq.~(\ref{eq:in:Riesz})].
For $k \to 0^+$, the Riesz gas converges to the log-gas that appears in random matrix theory. There, it is well known that the typical distribution of the gap in the bulk of the Wigner semi-circle, appropriately scaled, is well approximated by the Wigner surmise \cite{Mehtabook}, namely the gap between two particles in the case $N=2$. It emerged that from the studies in Ref. \cite{Santra2021} that this Wigner surmise does not however hold for $k\neq 0$. Our exact result in this paper, namely  ${\cal P}_{\rm gap, bulk}(g,N) \sim N H_\alpha(g\,N)$, indeed proves that the Wigner surmise does not hold for $k=-1$, corresponding to $k=1$. In this case, the Wigner surmise, using just two particles, would predict a gap distribution with a Gaussian tail, while our exact calculation of the scaling function $H_\alpha(z)$ shows that the gap distribution decays as $e^{-N^3\, g^3/(96 \alpha)}$, much faster than the Gaussian.

Concerning the second observable, namely the FCS, we have shown that the variance of $N_I$ is of order $O(1)$ and independent of $N$ and $L$ in the large $N$ limit. This is in stark contrast with the log-gas case $k \to 0^+$, where the variance behaves as $\ln(\bar{N}_I)$ where $\bar{N}_I \approx L \sqrt{N}$ for the log-gas with edges at $\pm \sqrt{2 N}$. Thus, compared to the log-gas, the fluctuations get suppressed further, making the $1d$OCP even more hyper-uniform. For the Riesz gas with general $k>-2$, several observables have been studied recently, such as the distribution of the position of the right-most particle \cite{KKKMMS21,KKKMMS22}. It would be  
interesting to extend these studies to FCS in the Riesz gas for other values of $k$, different from $k=0^+$ (log-gas) and $k=-1$ ($1d$OCP).    

\ack
We thank A. Dhar, M. Kulkarni and A. Kundu for useful exchanges.

\end{document}